\documentclass[a4paper]{article} 
\usepackage{geometry} 
\usepackage{fancyhdr} 
\usepackage{amsmath ,amsthm ,amssymb} 
\usepackage{graphicx} 
\usepackage{calc}
\usepackage[hyphens,spaces,obeyspaces]{url}
\usepackage[pdftex, colorlinks=true, allcolors=black]{hyperref} 
\usepackage{pdflscape}
\usepackage{enumitem}
\setlist[description]{labelindent=\leftmargin/2}
\usepackage[square,numbers]{natbib}

\title{Thoughts on Child Safety on Commodity Platforms}
\author{Dr Ian Levy\\Technical Director\\UK National Cyber Security Centre\\ian@ncsc.gov.uk \and Crispin Robinson \\Technical Director Cryptanalysis\\GCHQ}
\date{$21^{st}$ July, 2022}

\begin{document}

\maketitle

\begin{abstract}
The explosion of global social media and online communication platforms has changed how we interact with each other and as a society, bringing with it new security and privacy challenges. 
Like all technologies, these platforms can be abused and they are routinely used to attempt to cause harm at scale. One of the most significant offence types that is enabled by these platforms is child sexual abuse - both scaling existing abuse and enabling entirely new types of online-only abuse where the impacts on the victim are equally catastrophic. Many platforms invest significantly in combating this heinous crime on their platforms, referring confirmed evidence of illegality to law enforcement. The introduction of end-to-end encryption and similar technologies breaks many of the mitigations in place today and this has led to a debate around the apparent dichotomy of good child safety and good general user privacy and security. This debate has concentrated on the problem of detecting offenders sharing known abuse imagery using a technique known as client side scanning. 
We will show that the real problem of online child sexual abuse is much more complex than offender image sharing, providing a new set of `harm archetypes' to better group harms into categories that have similar technical characteristics and, as far as we are able, bring more clarity to the processes currently used by platforms and law enforcement in relation to child sexual abuse content and the real world impacts. We explore, at a high level, a variety of techniques that could be used as part of any potential solution and examine the benefits and disbenefits that may accrue in various use cases, and use a hypothetical service as an example of how various techniques could be brought together to provide both user privacy and security, while protecting child safety and enabling law enforcement action. 
 
We recognise that some of the techniques discussed require further research and work needs to be done to create a framework that can be used for consistent evaluation of various techniques on specific platforms and services.  These are highlighted at the end of the paper.
This paper is not intended to represent UK Government policy, but we hope it will lead to a balanced and informed debate that will help inform global policy in this area.

\end{abstract}
\section*{Executive Summary}
As more of our lives and economies have moved online, we increasingly rely on the plethora of global social media and communication services, and with that reliance comes certain privacy and security expectations from users. This expanded reliance also engenders an expectation user safety. As with any technology, these services are abused by malfeasants to cause harm to others, from cyber bullying to state sponsored disinformation, from cybercrime to online child sexual abuse. 
Child sexual abuse is a societal problem that was not created by the internet and combating it requires an all-of-society response. However, online activity uniquely allows offenders to scale their activities, but also enables entirely new online-only harms, the effects of which are just as catastrophic for the victims.

We hope this paper will help the debate around combating child sexual abuse on end-to-end encrypted services, for the first time setting out clearly the details and complexities of the problem (which is much more complex than other government needs, such as exceptional access) and looking to better frame the potential benefits and disbenefits of any solutions. We hope to show that the dual dystopian futures of safe spaces for child abusers and insecurity by default for all are neither necessary or inevitable. 
We have written this paper having spent many years combating child abuse, but also in the technical domains of cryptography and computer security. This paper is not UK government policy and is not intended to be a recipe for what governments could demand in the future. It is a genuine attempt to encourage debate and develop a common understanding of the problem, and the risks and benefits of any future technology changes. 

For many years, most mainstream social media and communication platforms have implemented technical mitigations that help protect some of their most vulnerable users, including children, from abuse taking place on the platforms and real-world abuse facilitated by the platforms. 
These technologies are used to detect potential child sexual abuse related activity which is then often referred to a human moderator to confirm illegal content, before being passed to the relevant national body that is authorised to deal with such referrals.
The majority of mainstream social media and communication platforms are US based, so this body is usually the National Centre for Missing and Exploited Children (NCMEC) via their `CyberTipline'.
 NCMEC reviews the content and, if appropriate, reports it to the relevant authority. In the UK this is the National Crime Agency (NCA).
In order to understand what mitigations are appropriate, it is important to understand the scale of online child sexual abuse. The statistic most often used to illustrate this is the number of reports received by NCMEC which amounted to 29.4 million in 2021. However, without context this number provides little useful information and can be easily misinterpreted. In the same year the NCA received 102,842 reports from NCMEC, but some of these were incomplete or, once investigated, not found to be child abuse. 
Of the 102,842 reports, 20,038 were referred to local police forces and started (or contributed to) investigations. In the same year, over 6,500 individuals were arrested or made voluntary attendances due to offences related to child abuse and over 8,700 children were safeguarded. 
These numbers more accurately illustrate the scale of the societal problem of child sexual abuse in the UK, of which the online component is significant. We would like to be able to show the causal link between individual CyberTips and convictions. However, this is not currently possible; industry notifications may lead to a completely new investigation, provide new evidence to allow investigations into an existing suspect or provide further evidence of scale of offending to an existing prosecution and we do not currently have the data to understand which of these outcomes has occurred in which cases.  However, we hope to be able to provide more in-depth analysis over the coming years as data collection improves.

Recently, many of these same platforms have started to remove their own access to user content, through technologies including end-to-end encryption, ostensibly to provide privacy for their users. This move fundamentally breaks most of the safety systems that protect users, and that law enforcement rely on to help find and prosecute offenders. As a result, governments around the world have vociferously raised the spectre of `safe places' where child abusers can operate with impunity, while academics and privacy campaigners have raised the spectre of a world of technology that is `insecure by default', where privacy and security are fundamentally impossible, with poor design choices justified through the exhortation `Think of the children!'. Both potential futures are possible. We believe that neither are inevitable or desirable. 

Countering child sexual abuse online is as complex in principle as exceptional access is simple. Our exceptional access essay said many times that `details matter'. In any analysis of how we counter abuse on encrypted platforms, there are many more details to consider. This isn't a unitary problem  - harms accrue in different ways and through different offender and victim behaviours. We believe that one of the challenges with this particular policy debate is that governments and law enforcement have never clearly laid out the totality of the problem being tackled. In the absence of that, people infer a model that turns out to be incomplete and, in some cases, incorrect. In publishing this paper we hope to correct that information asymmetry, and engender a more informed debate.
 
There are existing typologies and taxonomies of child sexual abuse, but none that focus on the specific ways in which the offences manifest online which define how they could be mitigated. We have created `harm archetypes' to try to frame the problem in a new way. These are: 
\begin{enumerate}
\item{Consensual peer-to-peer indecent image sharing\\
This where two children or young people voluntarily exchange nude or explicit images with each other. This is still the exchange of illegal imagery, but the outcomes resulting from the identification of this type of content must be very different to the other harm archetypes. }
\item{Viral image sharing\\
This is where people (usually) without a sexual interest in children share child sexual abuse images or videos in disgust or misplaced humour. These can go viral, causing significant further harm to the victims. This harm archetype seems to be the cause of many of the reports to NCMEC and so contributes significantly to the 29.4 million reports, but these are filtered before reaching NCA, so are not included in the 100 thousand referrals.}
\item{Offender to offender indecent image/video sharing\\
This is as described: offenders who are already in contact with each other, sharing illegal child abuse content. }
\item{Offender to victim grooming\\
Offenders will attempt to contact children online and convince them to meet in the real world (possibly leading to contact abuse) or send explicit, illegal images of themselves which often escalates to blackmail leading to more extreme demands from the offender. The second pathway of harm is relatively new, as it has been enabled by the commodity social media and communication platforms. It is worth noting that offenders will often use existing explicit images of children as a `trust token' with their victims, to try to convince the victim they're legitimate. }
\item{Offender to offender communication}
\item{Offender to offender group communication\\
Both archetypes involve offenders discovering and communicating with each other to normalise their behaviour, share tradecraft and techniques and even to plan real world abuse. We separate the archetypes because of the different technical characteristics for detection and mitigation. }
\item{Streaming of on-demand contact abuse\\
Offenders, usually Western, will pay to watch and direct the live abuse of children, often located elsewhere in the world. }
\end{enumerate}

Whether or not the above behaviours take place on a given online platform will depend on several factors, including:
\begin{itemize}
\item{the functionality of the platform}
\item{the technical capabilities of the offender}
\item{how potential victims interact with the platform}
\end{itemize}
For example, if a platform does not allow discovery of unknown users with particular characteristics, this is of limited use to an offender who wants to discover and contact children. By examining the technical characteristics and constraints of a platform, we can better understand how offenders exploit them and how to best to combat each harm. Even when similar harms exist on similar platforms, they rarely manifest identically, meaning that detection and mitigation must be platform specific. In our paper, we describe the harm archetypes in more detail and examine the service characteristics we believe are important in understanding the specific mechanisms of harm on a given platform, and therefore what may be necessary to combat that harm.

Many (but not all) service providers go to significant lengths to try to prevent, detect and refer child sexual abuse behaviour on their platforms. Many have access to the content of their users' messages, as well as significant volumes of complex metadata which is often used for targeting advertising on those services with that business model. This allows for relatively simple ways of combating child sexual abuse behaviour online, since:
\begin{itemize}
\item{images can be checked by server infrastructure using techniques like PhotoDNA (to ensure they are not known child abuse images)}
\item{text can be analysed for risky language and so on}
\end{itemize}
The metadata available to each platform provider varies significantly, with some holding vast swathes of complex metadata about each user and each interaction. Others hold almost nothing. Analysis of metadata may detect some types of behaviour which could suggest child sexual abuse. For example, a new account contacting many accounts which:
\begin{itemize}
\item{appear to be children, and}
\item{have a high rate of rejection}
\end{itemize}
may indicate initial grooming contact. But equally it could be spam. If service providers are willing to terminate these accounts quickly, then it can be argued that the harm is averted. However, offenders are often persistent and in most of the harm archetypes listed above, the risky behaviour can be harder to spot with confidence; for example, a child sending an explicit image to an offender. In these cases, more robust mitigations must be employed.

As service providers make design choices to include end-to-end encryption, the techniques in use today become less useful and some (such as servers checking whether images are known child sexual abuse material) simply cease to work. However, we do not think this rules out all opportunities for safe, private, comprehensive and effective systems that ensure user safety, and child safety in particular, on social media and communication platforms.
In this paper, we explore a range of techniques that could be employed to help reduce harm on a given platform where the service provider does not have access to user content. Again, this is intended to be a menu of potential mitigations, rather than a tick-list of things that must be implemented (since the technicalities of the platforms involved matter, as do the harm archetypes that will be prevalent on each one).
Researchers rightly point out that poor designs for safety systems could have catastrophic effects on user safety and security. However, we do not believe that the techniques necessary to provide user safety will inevitably lead to these outcomes.
For example, one of the approaches we propose is to have language models running entirely locally on the client to detect language associated with grooming. If the model suggests that a conversation is heading towards a risky outcome, the potential victim is warned and nudged to report the conversation for human moderation. Since the models can be tested and the user is involved in the provider's access to content, we do not believe this sort of approach attracts the same vulnerabilities as others.

That is not to suggest there aren't risks to this approach that wouldn't need to be mitigated. The system will need to be designed and built properly, but that is true of the rest of the service. The more subtle sociotechnical risks (for example the effect of a misclassification on the relationship between the parties) will need more research to be fully understood and mitigated. However, we believe that a robust evidence-based approach to this problem can lead to balanced solutions that ensure privacy and safety for all. We also believe that a framework for evaluation of the benefits and disbenefits is needed. We don't provide one in this paper, but note that REPHRAIN, the UK's national Research centre on Privacy, Harm Reduction and Adversarial Influence Online, is doing so as part of the UK government's Safety Tech Challenge Fund, although this will require interpretation in the context of national data protection laws and, in the UK, ICO guidance.

Intuitively, AI systems acting on the `behaviour' of accounts seems to be a solution to the problem at hand. AI systems can, in some circumstances, achieve highly accurate predictions and can work from data that will still be available once access to content is removed. However, in most of the harm archetypes, AI approaches that only use metadata are severely limited (with the possible exception of viral image sharing). We explore the various reasons why AI techniques alone are unlikely to be the solution to this problem, and also explore why access to verified illegal content is critical to law enforcement action. Many AI-based solutions being proposed do not give content to law enforcement, and instead use metadata about a user's account to establish likelihood of offending. This means that law enforcement agencies will be expected to act on a tip that basically says `Our AI says Person X is probably involved in something dodgy, but we can't explain why and we can't give you any evidence to back it up.' Any next steps that law enforcement could take - surveillance, arrest and so on - are highly intrusive and so have a high threshold for authorisation which this almost certainly wouldn't meet. Down this road lies the dystopian future depicted in the film {\em Minority Report}. 

In summer 2021, Apple released a feature called `NeuralHash' that sought to detect known child sexual abuse images, but running only on the user's device where the images are in clear, rather than on the company servers. This sort of technique is known as client side scanning and has received significant attention from industry, academic researchers and the media, even though it is only one of the many techniques we will need in the future to provide user safety at scale. Intuitively, the removal of the service provider's ability to detect known child sexual exploitation images on their servers can be mitigated by performing the scanning on the user's device or client. However, this relatively simple change (of where a technique is run) fundamentally changes the security and privacy properties of the approach, including how an adversary might exploit this change. As is usual, security researchers sought to understand the system and its potential weaknesses and published their results. We explore at length how these techniques could be made safe in this paper, but it is instructive to consider three key issues identified in those works, which broadly apply to any client side scanning technique.

The first is that it is relatively simple to create completely benign images that generate false positives and are identified as potential child sexual exploitation images. False positives are a problem with any image classification technique, and the real question is therefore 'what is the impact of an adversary exploiting this weakness?' The actual impact depends on the harm archetype being discussed. For example, offenders often send existing sexually explicit images of children to potential victims to try to engender trust (hoping that victims reciprocate by sending explicit images of themselves). In this case, there is no benefit whatsoever in an offender creating an image that is classified as child abuse material (but is not), since they are trying to affect the victim, not the system. This weakness could also be exploited by sending false positive images to a target, in order that they are somehow investigated or tracked. This is mitigated by the reality of how the moderation and reporting process works, with multiple independent checks before any referral to law enforcement.

The second issue is that there is no way of proving which images a client side scanning algorithm is seeking to detect, leaving the possibility of `mission creep' so that other types of image (those not related to child sexual abuse) are also detected. We believe this is relatively simple to fix through a small change to how the global child protection NGOs operate, such that we have a consistent list of known bad images, with cryptographic assurances ensuring that databases only contain child sexual abuse images which can be attested to publicly and audited privately. We believe these legitimate privacy concerns can be mitigated technically and it is likely that the legal and policy challenges are harder, but we believe they are soluble.

Finally, the issue of robustness is raised. That is, how easy is it for a motivated adversary to disable the detection on their device? Again, the impact of this varies between the harm archetypes. If used to combat `offender-to-offender image sharing' archetype, this is indeed an issue (since we expect offenders to used modified clients that do not report honestly), but it is not an issue in the `offender to victim grooming' archetype (since the potential victim is highly unlikely to use a modified application that has the detection disabled). Even in the first use case, there are mechanisms to make real-world exploitation harder. 

Through our research, we've found no reason why client side scanning techniques cannot be implemented safely in many of the situations one will encounter. That is not to say that more work is not needed, but that there are clear paths to implementation that would seem to have the requisite effectiveness, privacy and security properties.

Finally, responding to the challenge of `put your money where your mouth is', we provide an analysis of two of the harm archetypes on a hypothetical but realistic service. While there is certainly more work to do on the techniques we describe, including a more complete security and privacy analysis, we believe that these examples demonstrate that it should be possible to provide strong user safety protections while ensuring that privacy and security are maintained for all. We have suggested avenues of further work that we believe are necessary and hope that this paper motivates such work and a more inclusive and informed debate.

\section{Introduction}\label{introduction}

The issue of user safety - and in particular child safety - on commodity communication and social media services has been an issue since these services first became endemic. 
Over the years, many platform owners have worked with child safety organisations and national and international law enforcement agencies to help detect and combat this scourge on their platforms.
Put simply, these entities have worked together in order to 
\begin{itemize}
\item{Prevent online child sexual abuse\footnote{Child sexual abuse occurs online in a variety of different forms; in section~\ref{section:offencearchetypes} we set out an illustrative classification of the range of activities that are included under this general heading, and how they relate to harm to individuals and wider society.} wherever possible}
\item{Detect child sexual abuse where we have failed to prevent it}
\item{Work to investigate when such detections have occurred}
\item{Safeguard victims and bring offenders to justice.}
\end{itemize}

Of course, child sexual abuse is not confined to the online world, and this set of intents applies equally in child sexual abuse that has no online component.
This paper, however, concentrates on the issue of online child sexual abuse in its many forms and, in particular, how changes to commodity services around the introduction of end-to-end encryption (and other privacy and anonymity technologies) will affect the current mitigations. 

Robust detection of known child sexual abuse imagery is widespread on many services today, generally through computing server-side perceptual hashes on media (images and videos) sent over the service and comparing those to the perceptual hashes of known-bad media, curated by the relevant child safety organisations.
Techniques such as PhotoDNA, a long-standing perceptual hash that operates deterministically, are generally relatively robust to image manipulation and have excellent false positive rates, leading to a manageable\footnote{Though still significant, given the scale of the problem} workload for those who act upon these detections.
Server-side techniques running on the content of messages can also help detect other dangerous behaviours such as the grooming of children, which often leads to new image creation or real-world contact abuse. 
In each case, algorithms running on the service infrastructure today provide high quality information and content to human moderators, employed by the platform owners, who make the final decision about referral to the relevant child safety organisation.

More and more services are moving to end-to-end encryption of one sort or another with many espousing the privacy benefits this will bring to users\footnote{It is worth noting that much of the privacy intrusion brought by many of the {\em free at the point of use} services does not depend on access to content and so removing the ability of the platform to see content does little to impact the privacy intrusion generated by the platform itself that is monetised to pay for the service.}. 
It has been noted at length that adding end-to-end encryption will break many of the detection and mitigation techniques in use today, with a concomitant impact on real world child safety, but with little discussed about how that loss can be remedied. 
Consequently, child safety on end-to-end encrypted commodity communication services has become a totemic issue over the last year with highly polarised views, representing the challenge of balancing safety, security, privacy and access in the modern world. 

Much of the recent literature has concentrated on the issue of the detection of sharing of known child sexual abuse imagery on end-to-end encrypted services.
Sharing of existing, catalogued child sexual abuse images is a cause of significant harm to victims as they are revictimised every time an image is shared and it usually signals a wider sexual interest in children, which may require investigation. Viewing such imagery can further radicalise offenders and be a precursor to contact abuse.  
Conversely, there are situations when people who have no sexual interest in children whatsoever will share child sexual abuse images in outrage. While the interventions towards the people involved will obviously be different, the spread of these images must still be contained and instances removed since they revictimise the subjects and cause harm to unwitting recipients. 
However, sending images and videos of child sexual abuse is not the only harm related to child sexual abuse and concentrating on just this issue causes unhelpful bias in both examimation of the problem, evaluation of the utility of any proposed solutions and the relative benefits and disbenefits of any design on safety, security and privacy.
We believe that this previous narrow focus on image sharing in recent literature is influenced by a lack of accessible information around the problem of online child sexual abuse, as well as assumptions about what sort of approaches may be necessary in future to create safe environments for vulnerable users, to ensure it is possible for platforms to minimize the incidence of child sexual abuse on their services and for law enforcement to identify and investigate serious offences, safeguarding the victims where possible. 

This paper attempts to address those issues.
We will define fully the problem of online child sexual abuse and describe, as far as is possible in a public paper, offender behaviour that is relevant to designing and evaluating technical mitigations. 
We will describe the different archetypes of harm related to child sexual abuse online and how they come to pass. 
We will describe the characteristics of the current set of mitigations and how they are affected when they cannot access content. 
We will suggest some putative solutions that may be of use in end-to-end enrypted scenarios and then give a very high level suggestion of how a set of reasonable\footnote{In the authors' opinions.} technical mitigations can be applied, along with our view of the evaluation of the benefits and disbenefits.

\subsection{What this paper is}

This paper is intended to provide a basis for a better public debate about safety systems\footnote{We concentrate on child sexual abuse in this paper not to evoke a particular emotional response in the reader, but because it is high harm, international in nature and there is almost universal intolerance.} on modern commodity services, and the potential impact on user safety, privacy and security.
The authors have been involved in providing technical support to the UK community countering child sexual abuse online for many years, and so have some experience in this area. We have also consulted many involved in detection, prevention and management of child sexual abuse in the UK across government, law enforcement and the relevant child safety charities.
Any errors or poor logic are the responsibility of the authors alone.
This paper is intended to be neutral.
We are attempting to educate and inform in order to allow a more detailed and informed debate to happen; we are not attempting to stifle that debate. 
We have tried to be neutral in our descriptions and analysis; where we have failed, it is accidental. 

\subsection{What this paper is {\em not}}

\begin{itemize}
\item{This paper is not a high level design document.}
\item{This paper is not a full security analysis of any particular solution.}
\item{This paper is not a legal or privacy analysis of any particular solution.}
\item{This paper is not a set of requirements that the UK Government wishes to be imposed on commodity services.}
\item{This paper is not intended to be an exposition of the entirety of public safety concerns on commodity systems, but is limited only to child safety issues.}
\item{This paper is not an exposition of UK Government policy, nor are any implications that can be read in this document intended to relate to UK Government future policy\footnote{They are likely just bad drafting by the authors.}.}
\item{This paper does not seek to reopen debates on exceptional access.  While the six principles in our Lawfare article on exceptional access \cite{lawfare2018} apply to any safety system, there is no implied read-across in terms of technical or policy implications for exceptional access.}
\item{This paper should be proof that {\em details matter} when talking about this subject. Discussing the subject in generalities, using ambiguous language or hyperbole will almost certainly lead to the wrong outcome.}
\end{itemize}

\section{The Child Sexual Abuse Problem Online}\label{offenders}

There is a significant information asymmetry in this space.
Many who are concerned about the potential privacy and security impacts of user safety systems are not inculcated in the online child sexual abuse problem, often working from incomplete data or assumptions.
In this section we try to address that lack of information.
We explain the scale of child sexual abuse online in relation to the UK, decompose the generic problem into archetypes with different characteristics and describe, to some degree, how protections currently function.
Later sections of this paper discuss implications of modern privacy technologies, such as end-to-end encryption, and offer thoughts on potential solutions.

It would be unhelpful in the wider sense to disclose precise details of how offenders operate and what specific characteristics currently help law enforcement detect and prosecute them. Offenders constantly update their tactics to circumvent platform protections and law enforcement and  
this will limit the detail of what is in this paper, but we have tried to ensure that there is sufficient information to inform the debate.

\subsection{Harm to Victims}
Here, we reproduce the overarching description of harm caused by child sexual abuse from the government's Tackling Child Sexual Abuse Strategy \cite{csea-strategy}. 

\begin{quotation}
Whilst the impact of child sexual abuse on victims and survivors can vary significantly, there is strong evidence that child sexual abuse is associated with an increased risk of
adverse outcomes in many areas of a person's life. This can include physical, emotional
and mental wellbeing, relationships, socioeconomic outcomes, and vulnerability to revictimisation. The impact of child sexual abuse can be significant, regardless of the type
of abuse suffered (including where the abuse takes place in an online environment), and can be influenced by a range of factors including the duration of the abuse, an individual's
coping mechanism, and the support they receive.

Research by the Independent Inquiry into Child Sexual Abuse (IICSA) demonstrates that the impacts of child sexual abuse can last for a
lifetime, sometimes resulting in long-term illness and disabilities. These can include a wide
range of physical health conditions, as well as mental health issues such as depression,
anxiety disorders, and post-traumatic stress disorder (PTSD). Rates of self-harm have been
shown to be as high as 49\% among adult victims and survivors in treatment, and the risk of
victims and survivors of child sexual abuse attempting suicide can be as much as six times
higher than the general population.

The IICSA's rapid evidence assessment on the impacts of child sexual abuse~\cite{iicsa-rea} highlights the following key impacts of child
sexual abuse:
\begin{itemize}
\item{Physical health\\
Physical injuries, high BMI, problems related to childbirth, unexplained medical problems}
\item{Emotional wellbeing, mental health and internalising behaviours\\
Emotional distress, trauma/PTSD, anxiety, depression}
\item{Externalising behaviours\\
Substance abuse, `risky' and inappropriate sexual behaviours, offending}
\item{Interpersonal relationships\\
Reduced relationship satisfaction, issues with intimacy and parent-child relationships}
\item{Socio-economic\\
Lower educational attainment, higher unemployment, financial instability, homelessness}
\item{Religious and spiritual belief\\
Disillusionment with religion, faith as a coping mechanism}
\item{Vulnerability to revictimisation\\
Sexual revictimisation in childhood and adulthood, other types of victimisation}
\end{itemize}

Victims and survivors may face barriers to progression in education and to their careers, with research suggesting, on average, victims and survivors have higher rates of unemployment and long-term sickness. They may also use negative coping mechanisms to deal with the impact of abuse. Research points to higher rates of substance misuse amongst victims and survivors of child sexual abuse.

Experiencing child sexual abuse can also increase the likelihood of further victimisation.  The 2018-2019 Crime Survey for England and Wales has shown that those who experienced child sexual abuse were significantly more likely to experience domestic abuse and further sexual abuse as adults.

It is also important to acknowledge that child sexual abuse and subsequent criminal investigations can have a profound impact on the family members of both victims and perpetrators, including social, psychological and financial consequences. 
The significant impact on alleged offenders may accrue when {\em accused} of involvement in child sexual abuse, even if that accusation turns out to be incorrect and investigations must be cognisant of that. 
There is a need to develop our understanding of the impact of these crimes and investigations on the wider family, and work is underway with law enforcement partners and voluntary sector organisations to explore these issues further. 
\end{quotation}

This extract cannot do justice to the problem and readers are encouraged to explore the expert child safety literature available including, but not limited to \cite{csa-typology} \cite{canadian-survivors-report} \cite{phoenix11-statement} \cite{IICSA-report-2020} \cite{LFF-report} \cite{iicsa-rea} \cite{ev-philippines}.

\subsection{The Scale of the Problem}\label{section:scaleofproblem}

In order to assess the significance of a societal issue like online child sexual abuse, and therefore what an appropriate response may be as the threat changes, it is necessary to understand the scale and severity of the problem.

Firstly, it is important to understand the process that enables reporting and investigative work today on many, but not all, commodity services.
This consists of the following steps: 

\begin{itemize}

\item{All major platforms have a content moderation policy (sometimes referred to as Community Standards) which include clear prohibitions on activity, behaviour and content related to child sexual abuse.
Users of the platform formally accept these conditions as part of the terms of service when they sign up.}

\item{The platforms deploy various technical measures, including techniques to detect known child sexual abuse material and behavioural techniques to identify risky behaviour, to identify this type of activity.
Activity highlighted as potentially being in contravention of the content moderation policy, including any user reporting, is usually reviewed by a human moderator.
If confirmed as child sexual abuse related content, there is then a legal obligation on the platform, under US and other national legislation, to report the incident to the designated national organisation.
For US companies it is the National Center for Missing and Exploited Children (NCMEC).
These referrals are reported to NCMEC's CyberTipline and are called `CyberTips'. The Online Safety Bill seeks to introduce a similar UK reporting requirement.} 

\item{Substantial numbers of CyberTips are sent to these clearing house organisations around the world.
CyberTips from the main commodity platforms are broadly reported to NCMEC since the largest social media and communication platforms are based in the US.
They review the tips, adding other information they hold and then sending a composite CyberTip to the relevant authority in the country where the activity has originated.
For the UK they are sent to the National Crime Agency (NCA).} 

\item{We can only speak for the UK regarding follow-on activities.
NCA review every single report they receive to determine if there is sufficient information to launch an investigation or support an existing case. All referrals received are retained by NCA for intelligence purposes.}

\item{Should there be sufficient information to allow a law enforcement intervention, a range of investigative options will be assessed and pursued as appropriate, including the potential for applications for warrants and arrest. Importantly, interventions may be made to safeguard the victim or victims involved. 
The normal judicial process is followed from this point and is not enumerated further here.}

\end{itemize}

It is worth noting that the content moderation policies are not legally mandated  - although the UK's Online Safety Bill seeks to bring in such a duty - but are put in place by the platforms for a combination of commercial,  moral and ethical reasons. 

Understanding the scale of the problem and the harm is difficult, not least because reporting periods, thresholds for report or action, normalisation and so on are not common across the data sources that must be combined. Furthermore, there are limits on the authors' ability to access detailed reports (as there should be!), making data analysis more difficult.
Nevertheless, we will attempt to present an unbiased picture of the data we can obtain. Again, these data are UK specific, although we expect similar patterns to be seen in other Western countries.

In the calendar year 2020, NCMEC received 21.75 million CyberTips from US headquartered platforms. In calendar year 2021, this number was 29.4 million with those reports covering a total of more than 80 million images and videos\footnote{Note that we do not suggest that this represents 80 million distinct items of child sexual abuse content, rather 80 million occurrences of child sexual abuse images and videos being shared online.}. 
This is an important number from the perspective of understanding the scale of the problem that the platforms face, noting that it only represents the activities that are identified rather than the totality of the offences being committed.
That number includes activities that go across all the archetypes of harm described later in this section, and so it is questionable as to whether this number alone is a useful proxy for the harm caused by child sexual abuse activity on US platforms. 

For the calendar year 2021, NCA received 102,842 UK related industry reports concerning child sexual abuse material, up 24\% on the previous calendar year. 
These reports will include informational reports, which are those cases NCMEC has had reported to them by internet companies, which NCMEC or the company has predetermined to be either viral images, adult pornography, children who are clothed, anime or cartoon, or where the  company has acknowledged an omission due to a technical reason (normally images are missing).
Informational reports are filtered from further action as they either do not contain sufficient information to initiate an investigation, or do not contain child sexual abuse content. In calendar year 2021, the NCA referrals bureau made 20,038 disseminations of reports to local forces and NCA teams, compared with 16,651 in 2020.
 These reports all started or contributed to investigations. In calendar year 2021, NCA state that UK law enforcement made over 6,500 arrests or voluntary attendances and safeguarded or protected over 8,700 children as a consequence of industry reports of child sexual abuse. 
It would be useful to be able to show that some number of CyberTips uniquely provided evidence for some number of convictions and safeguarding events. Due to the multi-modal nature of offending and the lack of consistency in data recording, this is currently not possible. Industry notifications may lead to a completely new investigation, provide new evidence to allow investigations into an existing suspect or provide further evidence of scale of offending to an existing prosecution case. 
We hope to be able to provide a more in-depth analysis over the coming years as data collection is improved. 

A `voluntary attendance' is the process by which a suspect is interviewed in relation to an investigation  under caution whilst not under arrest. The suspect still has a right to legal representation.
Voluntary attendance can be used as an alternative to arresting a suspect within an investigation when the necessity conditions for arrest, as set out in the Police and Criminal Evidence Act \href{https://assets.publishing.service.gov.uk/government/uploads/system/uploads/attachment\_data/file/903814/pace-code-g-2012.pdf}{Code G}\footnote{https://assets.publishing.service.gov.uk/government/uploads/system/uploads/attachment\_data/file/903814/pace-code-g-2012.pdf}, are not met. It can be the case that these conditions are not always met in a child sexual abuse investigation, or any investigation for that matter, and this allows the subject to be interviewed whilst not being detained (i.e. they are present voluntarily).  
In the context of industry referrals leading to law enforcement action for child sexual abuse investigations, voluntary attendance can and should be used in addition to arrest figures to give an accurate picture of the response. In both circumstances a referral has led to a suspect being identified and interviewed. Crucially, both voluntary attendance and arrest can result in an individual being charged and prosecuted. 

A `safeguarding action' can include taking action against those in the children's environment, for example professionals, volunteers, foster carers and so on. Such action is normally determined at the Local Authority level and may include referral to law enforcement, but other agencies may undertake safeguarding actions. 

`Protect actions' refer to an individual child at risk of, or suffering, significant harm. This threshold is usually determined by statutory action undertaken under Section 47 of the Children's Act 1989. Outcomes could include removing the child into care, or other significant interventions to stop further harm. 

Another action that can be taken is education of the children involved in `sexting' as to the risks they are taking. We use the data published for law enforcement outcomes in 2021, published in \cite{outcome-data-table} to illustrate scale. 
Indecent images of children (which is how self-generated images would be categorised, as well as abuse images) are covered by the `Obscene Publication' category. While this category does not uniquely consist of offences involving indecent images of children, the majority of cases do and both Home Office and law enforcement believe this to be a good proxy metric. 
In the year ending 31st March 2021, there were 7,312 Obscene Publication offences given an outcome of `Further investigation to support formal action not in the public interest - police decision', which is known as `Outcome 21'. Understanding the Outcome 21 statistics is more complex, as individual offence types are not published and application of this outcome is at the discretion of the local force. We note that the recording process and the fact that discretion is applied at local force level means that there will be error in this number. However, it is a useful order of magnitude measure.

While some of the statistics used in public do not accurately depict the scale of the harm, these real-world outcomes do demonstrate the scale of the harm that is mitigated by the current efforts of platforms to tackle child sexual abuse.
While there is complexity, error and subtlety in these data, we believe that the data presented here show that online child sexual abuse relevant to the UK occurs at a significant volume on commodity platforms and warrants significant investment in mitigation efforts. It should also be obvious that the data we have are likely to represent only some percentage of the actual volume of illegal child sexual abuse activity on commodity platforms.  
The UK is but one medium-sized country and proportional figures for arrests, voluntary attendances and child safeguarding are seen across numerous other developed nations.

\subsection{Offence Archetypes}\label{section:offencearchetypes}

In order to assess any technical mitigation in this space, we first need to be clear about what the technique is aiming to achieve, which in large part equates to defining the behaviour or offence that it is intended to prevent, detect or mitigate.
What follows is a high-level description of the primary online child sexual abuse offence archetypes  we consider in our analysis.
We will return to this again later in the paper, after we have reviewed the potential technical mitigations.

It is worth noting that this set of archetypes is focused on harm with an online element and categorised by technical characteristics.
For a more general typology of child sexual abuse see, for example, \cite{csa-typology}.
In this section we are focussed on the harms themselves rather than the measures (technical, procedural and societal) that are used to prevent and detect them; we offer a fuller presentation of such measures as employed today in section~\ref{section:current}.

\subsubsection{Consensual peer-to-peer indecent image sharing by a child}

This is also referred to as first person produced imagery and commonly known as `sexting'.
A child will take an indecent image of themselves and send to a peer\footnote{In this case, an online identity that really is a peer, rather than an offender masquerading as a peer, which we treat separately.}, as part of their social development.
Even when there is no malintent, characterised by coercion, deception or exploitation of the child involved, the creation and transmission of the image is a criminal offence.
However, the critical aim here is one of education and harm reduction where there is no evidence of coercion, deception or exploitation. In particular, the primary aim in this case is not prosecution.

Regardless of whether the recipient of the image was acting in good faith, if the child regrets the image, they can report it to the appropriate charities and NGOs, for example the IWF and NSPCC `Report Remove' service\footnote{The `Report Remove' service is at https://www.iwf.org.uk/our-technology/report-remove/}.
Our primary aim in this situation is to reduce further promulgation of the image.
This is to try to protect the subject, to stop the image being used in grooming scenarios to bolster a fake persona and to stop the image being shared by individuals with a sexual interest in children.
The image is often added to an NGO-run database and, in the UK often also the Child Abuse Image Database\footnote{A separate database of known child abuse images operated by and for UK law enforcement agencies.}, of known indecent images of children and further dissemination is blocked through on-platform scanning.
Issues with user reporting are explored briefly in section~\ref{subsection:user reporting}.

There is an apparent conundrum in that the child's actions constitute a criminal offence, but the intent is to not criminalise them.
In the UK, law enforcement use `Outcome 21' which enables them to deal with such offences but without criminalising the child.
This allows a proportionate response to the offence and the opportunity to ensure that no coercion, deception or exploitation was involved which may itself require a law enforcement investigation.

For more information about the use of Outcome 21 in this situation, see \cite{outcome21-report}.
It should be obvious that Outcome 21 is unlikely to be appropriate in relation to any further sharing or dissemination of the image outside the original relationship and, in some cases, a small peer group.

\subsubsection{Viral Image Sharing}

This relates to actual child sexual abuse imagery, rather than cartoon or generated imagery, being shared in a meme or viral-like fashion.
Often the majority of people sharing this imagery do not have a sexual interest in children and are sharing in outrage or even misplaced humour. Regardless, dissemination remains an offence. 

However, this activity is far from harmless and causes significant, ongoing trauma to the victims depicted in the images. Often, it is distressing for the unwitting recipients as well. 
A recent example involved a Category A video, the most serious category, of a male child aged 2 or 3 being sexually abused which was shared at scale in a viral manner on a social media platform.
We believe that these viral sharing events produced of the order of 3 million NCMEC reports, mainly driven by people who were outraged, trying to help identify and rescue the victim. 
This action causes lifelong harm to the victim and contributes to the normalization of this type of material and the abuse that it shows in the minds of those with a sexual interest in children.
Detection and removal of this sort of sharing can be very beneficial to all users involved, in many cases, and that benefit is increased the earlier in the dissemination cycle it can be achieved (with the best outcome being that the image is never received by anyone). 

However, offenders also contribute to this category, hoping to hide in the noise as they reshare the material with like-minded individuals.
 As a consequence NCMEC referrals of material in this archetype regularly lead to the identification and prosecution of even more serious offences.
There is significant complexity in determining which parties, if any, in a given viral sharing event have a sexual interest in children, and so need to be referred to the authorities, and which are not, and so need to be educated.

\subsubsection{Offender to offender indecent image/video sharing}\label{section:imagesharing}

The description of this category is self-explanatory - offenders who are already in contact with each other exchanging illegal content.
In many services today, perceptual or fuzzy image matching algorithms, such as PhotoDNA, are used to detect the transmission of child sexual abuse imagery\footnote{Imagery and video, but the issues are the same so we talk about imagery only for brevity.} that is already known to and validated by one of the worldwide child protection NGOs.
These are also colloquially known as `hashes' but have very different properties to traditional cryptographic hashes.
Some services also seek to detect new, previously unseen child sexual abuse material that is not already held in the relevant databases. This is obviously a hard problem with higher false positive rate than deterministic hashes of known images and often contextual information is used - as well as image content - to try to help make a determination. 

Today, many, but not all platforms, scan imagery and video as it is uploaded or sent to determine whether it is flagged as illegal child sexual abuse material.
If the content is flagged, it is usually referred to a human moderator, employed or engaged by the company that owns and runs the service\footnote{Some companies choose to have specially trained and supported moderators who verify potential child sexual abuse content, while others choose to refer directly to the relevant child protection NGO.}.
Only if the human moderator concurs that the content is illegal is it referred to one of the global child protection NGOs (for US-based services, this is NCMEC) who verify illegality and then notify the relevant law enforcement agency in the jurisdiction in which the offender appears to be based. It is worth noting that these specific NGOs must have legal authority to perform this role in their home jurisdiction, otherwise they are themselves {\em a priori} breaking the law. 
 
It is hard to overestimate the harm caused by this activity.
The scale and patterns of online abuse raises questions around the potential for normalisation of this kind of behaviour among offenders, including through discussions with like-minded individuals on internet forums, and the potential for escalation into contact sexual abuse of children \cite{csea-strategy}. 
Recent research \cite{risk-factors} suggests that access to child sexual abuse material online causes some consumers of this material to then seek direct contact online with children. However, this work is based on a self-report survey and so is somewhat limited and not wholly representative, but nonetheless suggests that it is important to limit access to child sexual abuse material online.
It also revictimises the children involved with every share or trade causing them further harm.
Many victims report that they are scared of being identified later in life and all further distribution of the content increases their risk and anxiety.
There are a multitude of examples of victims being identified by offenders in later life, both online and in person.
It is difficult to imagine the trauma felt by a victim of child abuse when they are recognised and approached in real life by someone with a sexual interest in children. 
The Canadian Centre for Child Protection `{\em surveyed abuse survivors and observed that 
a significant proportion of respondents (69\%, n=129) said they worried about being recognized by someone who had seen the imagery of their abuse and described how that worry impacted their daily lives. That worry was justified - 30\% (n=99) said that they had been identified online or in person by someone who had seen the imagery, and some who had been targeted as a result (23 respondents - most of whom are still under the age of 40).}' \cite{canadian-survivors-report}.

The `Phoenix 11' group of child sexual abuse survivors have also talked about the effect on them, stating `{\em  We shared openly about being sexually abused as children, having that abuse recorded, and facing the constant fear that there was widespread and continuous distribution of the most horrific moments of our lives.}' \cite{phoenix11-statement}. 

\subsubsection{Offender to Victim Grooming}\label{section:grooming}
This class of harm involves direct communication between one or more offenders and potential victims.
In most cases, the offender will be acting under a persona that is likely to engender trust in the potential victim, normally that of a child. That persona may be present across a number of different social media and messaging platforms to further bolster it.

There are three main pathways to consider here: 
\begin{enumerate}[label=\alph*]
\item{Offender sending explicit images of children to the potential victim in order to bolster the persona and build trust.\label{section:offenderimageexchange}}
\item{Offender causing the victim to create new content, which leads to blackmail and escalation of abuse.
The victim is forced to create ever more extreme content, often including siblings and peers.}\label{section:offenderblackmail}
\item{Offender causing the victim to create new content, leading to physical meeting and contact abuse (possibly, but not always, via the previous pathway).}
\end{enumerate}
It is obvious that the harm to victims is severe.
 
Some offenders create and develop multiple online personas over extended periods of time, building up a realistic history of interaction with other users and exhibiting the appropriate behaviours for the age and interests associated with their persona.
Services which offer user discovery and which are popular with minors are particularly attractive to offenders, since they give them a mechanism to identify, contact and communicate with potential victims.
Once trust is established in the eyes of the victim, offenders will often move further communication to other messaging platforms that they feel are less likely to be monitored effectively. 
The National Crime Agency Operation Makedom, which led to the prosecution of Abdul Elahi, shows this behaviour in context. The press release\footnote{\href{https://www.nationalcrimeagency.gov.uk/news/sadistic-blackmailer-and-paedophile-jailed-for-32-years-after-targeting-nearly-2-000-people-worldwide-to-commit-sickening-online-sexual-offences}{https://www.nationalcrimeagency.gov.uk/news/sadistic-blackmailer-and-paedophile-jailed-for-32-years-after-targeting-nearly-2-000-people-worldwide-to-commit-sickening-online-sexual-offences}} details the scale of his offending and states that {\em `As soon as possible, Elahi moved victims onto WhatsApp - which is protected by end-to-end encryption - and away from the websites he met them on'}.

\subsubsection{Offender-to-offender communication} 

This category of harm is related to offenders discovering each other and engaging in one-to-one communication.
They will typically seek to reassure each other that their behaviour is normal, further normalising child sexual abuse in their minds.
This can also lead to escalation of offender behaviour, including incitement to create first generation imagery in order to gain entry to restricted access child sexual abuse communities and sites.
Many of these communities require the presentation of new child sexual abuse content in order to prove credentials which, as a side effect, often causes the groups to have the effect of validating increasingly serious behaviour (both on entry, since everyone in the group has had to go through the same process of producing new material, but also in an enduring way since they act as an echo chamber).

Offenders also share tips on which platforms to use, how to avoid detection when sharing child sexual abuse material and when grooming children on various platforms and techniques that make tracking them harder.

\subsubsection{Offender-to-offender group communication}

This is similar in nature to the previous category, although from an investigative perspective there are some additional opportunities to gain access or insight into the communications.

We have enumerated this separately because technical solutions are likely to be very different to the previous category.

\subsubsection{Streaming of on-demand contact abuse}\label{section:livestream}

This typically involves contact abuse taking place overseas, although it does also occur in the UK\footnote{See, for example, \href{https://www.wired-gov.net/wg/news.nsf/articles/Woman+admits+livestreaming+child+sexual+abuse+11112019141500?open}{https://www.wired-gov.net/wg/news.nsf/articles/Woman+admits+\hspace{3pt}livestreaming+child+sexual+abuse+11112019141500?open}.}, but which is being viewed in real time, and is usually directed, by an offender in (for example) the UK.
This is normally a business transaction where the person performing the physical abuse is paid by the offender.
Often, the abuser is a family member of the victim.
The impact on the victim is severe and the activity further normalises the offending behaviour.
The sites that offer the procurement of this service are normally traded between offenders, either bilaterally or in trusted groups.
There is some advertising of these services on social media sites and furthermore money does change hands, admitting some investigative routes, depending on the precise mechanism for payment (on-platform, off-platform or cryptocurrency). 

There is a further modality whereby individuals who advertise sex shows on (legal) adult streaming services and social media platforms are then asked to commit abuse on children for more money, with a proportion going on to do so.

\subsection{Importance of content for Law Enforcement Response}

Law Enforcement Agencies need reliable intelligence that an offence is likely to have been committed before they are able to take actions to safeguard children or to investigate suspected offenders.
The level of evidence required to obtain warrants for such action in the UK is very high and this is proportionate since the intrusion into the private lives of these suspects is substantial. 
Furthermore, to be falsely accused could cause significant harm to the individuals involved.
At present, the content itself is required to instigate real-world investigations stemming from online tips.

Content provides law enforcement with :
\begin{itemize}
	\item{Evidence of motive.}
	\item{Evidence of planning.}
	\item{Evidence of timescale to execution of the crime.}
	\item{Evidence of conspiracy with others.}
	\item{Evidence of the extent of an individual's role and culpability.}
	\item{Evidence of the extent of the crime, for example the number of victims and number of offences.}
	\item{Identification of historic, current and future victims.}
	\item{Identification of exculpatory material or confirmation of alibis.}
\end{itemize}

Public narratives have been presented~\cite{facebook-position-on-eprivacy},~\cite{cdt-outside} that suggest that metadata alone, processed by machine learning or artificial intelligence techniques would be sufficient to allow law enforcement to intervene. As we will explore later in this paper, these techniques will produce some probability that an account has (for example) engaged in grooming children. It is hard to envisage how such an algorithmic probability of malicious activity with little supporting evidence could be used to convince a judge\footnote{In the UK, there are multiple authorization regimes that may have different authorizing parties, for example a Secretary of State and Judicial Commissioner, a judge or a senior law enforcement officer. We use `judge' as shorthand.} that the investigation was necessary.

There have been suggestions in the public debate that it is the responsibility of law enforcement to obtain sufficient evidence given a probabilistic tip. This does not stand up to even light scrutiny. 
Consider the case where a probabilistic tip is given to law enforcement along with accessible service metadata.
Further assume this is sufficient to convince a judge or, dependent on the activity planned, the relevant authorising officers within law enforcement, to authorise further evidence gathering processes to be instigated. 
These processes include a range of covert and overt options that are, by their very nature, some of the most intrusive powers available to law enforcement. Examples include, but are not limited to, the arrest of an individual and their detention in custody, the physical surveillance of an individual or the search of private property under a warrant.   

In all cases, the intrusion is significantly higher than a platform moderator having access to specific portions of specific conversations for specific purposes on that platform\footnote{In making this statement, we assume that such moderation on an end-to-end encrypted service is granted only when specific tests are met, as we will explore later in this paper.}.
In the case of attacking the device, if law enforcement's only mechanism to access intelligence and evidence is the use of zero-day vulnerabilities in commodity devices, then this will drive a market in such vulnerabilities, reducing the number that are reported to vendors and making everyone less safe.
This is a real impact on cybersecurity of getting user safety designs wrong, as we discussed previously in our Lawfare blog \cite{lawfare2018}. It should be obvious, but it is worth stating that neither responsible law enforcement or responsible governments wish to engender such global vulnerabilities in commodity services or devices.  While this paper concentrates on countering child sexual abuse, the duty of governments and law enforcement is to protect citizens from all criminal and national security harms.

\subsection{Offender Hierarchy}

The National Assessment Centre's latest \href{https://www.nationalcrimeagency.gov.uk/who-we-are/publications/533-national-strategic-assessment-of-serious-and-organised-crime-2021/file}{National Strategic Assessment of Serious and Organised Crime}\cite{nca-2021report} reveals that there are estimated to be between 550,000 and 850,000 people in the UK with varying degrees of sexual interest in children who pose a concomitant level of risk to children.

Obviously, not all of these people with a sexual interest in children are engaged in contact abuse and it is helpful to try to categorise offenders in some way in order to understand their motivation and to determine appropriate responses.
If we consider classes of offender categorised by the strength of their sexual interest in children coupled with their ability to act on it, then each more extreme category will contain fewer people.
Splitting this population into four classes is probably sufficient for our purposes, although it is worthy of note that people in the more extreme categories will usually also engage in activity in lower categories. Readers should not interpret this as a quantized set of behaviours, but a continuum through which offenders will move over time, recognising that this may not be linear and that some contact offenders will not have been involved in online offending. 

\begin{enumerate}
\item{The first category is those who are just starting to explore their sexual interest in children.
They will mainly seek to find content in relatively accessible groups and services, often using clear web searches\footnote{Commodity web searches remain able to find such content, despite the current efforts of the service owners, child safety NGOs and law enforcement. See for example \cite[paragraph 88]{IICSA-report-2020}.} to discover imagery and groups of like-minded people.
Some of them are likely to be dissuaded from further action if they are `nudged' towards help services at the appropriate time, but law enforcement continues to see a level of recidivism in offenders and so this will always have limited effect.
UK law enforcement and children's charities invest effort in making help available and signposting that help where possible, for example when people search for common child sexual abuse related terms.
As a general principle, these people have low sophistication operational security, but some find or are directed towards instructions that will make them less likely to be detected.}

\item{The second category is those who have progressed to accessing specific child sexual abuse services and content that are more difficult to access, for example, those hosted on Tor Hidden Services or dedicated child abuse forums on open platforms that require new members to be vouched for by existing members.
These people will be technically more savvy, but sometimes will not be directly involved in child sexual abuse.
These people believe that their behaviour is - to some degree - normal.}

\item{The third category is those who have progressed to causing abuse, for example as per our harm archetypes in \ref{section:grooming}\ref{section:offenderblackmail} and \ref{section:livestream}.
They will be active in the child sexual abuse community, running groups and services.
They are technically savvy and understand the broad mechanisms used to discover child sexual abuse material and activity.
They have been involved in child sexual abuse for some time and believe their behaviour is normal.}

\item{The final category is those who have progressed to engaging in contact abuse.
When arrested, offenders of this type rarely show remorse or believe that their behaviour is wrong.}
\end{enumerate}

It should be obvious from this description that a variety of very different interventions are necessary for each group of offenders and their likely activities, and not all will apply to commodity communication and social media platforms.
The interventions and technologies we describe in this paper are intended to cover those activities that occur on commodity communication and social media platforms.
The types of mitigation and intervention for other types of service and communication - for example, Tor Hidden Services, offender communities that use services run by that community and so on - are out of scope of this paper, but are amenable to more direct intervention by law enforcement and, in the case of the UK, GCHQ.
For the purposes of this discussion, readers should assume that other appropriate national and international capabilities are brought to bear against these non-commodity technologies.

\section{Service Characteristics}\label{section:servicecharacteristics}

As is the case in understanding any complex system and the impacts of changes to it, details matter in this debate.

A diverse range of communication services are available today, with a commensurate diversity in features that affect the risk of the various different types of child sexual abuse occurring.
These differences impact the safety of users of the services, suggesting that the proportionate response to the child sexual abuse threat will have different characteristics on each service.

We shall enumerate the key characteristics of communication services that we believe are relevant to the evaluation of risk of child sexual abuse on commodity services.
We do not claim that the following enumeration is either exhaustive or normative.

\subsection{Peer discoverability}
Some services actively seek to encourage discovery of peers, for example if they set out to connect users with previously unknown people, or to allow them to discover online personas of people they might know or share similar interests with.
On others, a user needs to know some semi-private information (such as a phone number or email address) of a peer in order to connect with them.
This distinction has obvious implications for grooming and for offenders being able to identify potential victims, though it may be possible for services that usually offer peer discovery services to partially mitigate this effect by restricting their availability and reach to a subset of users, for example not allowing (asserted) adults to contact (asserted) children.
It should be noted that even when services do not themselves actively support discovery of peers, there often exist third party services that provide similar functionality, though these may only provide access to the peers that have registered with them (or whose account details have been published).

\subsection{Peer contactability}

Related to the above, whether or not a user is able to interact with peers with whom they do not have an established mutual trust relationship affects the ability of offenders to interact with those they have identified as potential victims.
It is worthy of note that interoperability between social media, messaging and other services may cause further confusion in this area, since multiple identities, behaviours and other account characteristics may be interpreted by the system or rendered to the user in different ways. 

\subsection{Connection to public services}
Some messaging services are a part of a wider product ecosystem including elements that are published either to the general public, or to a restricted audience in the control of the user or service provider (and visible to the servers).
This may provide a service provider with additional data with which to assess the threat to children that a user might present.
It is, however, worth noting that - just like in the real world - we cannot reliably expect behaviour in public to be indicative of behaviour in private, which limits the effectiveness of such techniques.

\subsection{Metadata available to server}
Non-content data, such as message size, send time/date, country of origin, and image type, is likely to be available to servers.
Some services allow the server to reliably identify the sender of a message in order to facilitate the communication, others have deliberately removed this ability and only know the identity of the recipient in order to successfully route the message to its destination.
Such restrictions reduce the effectiveness of any metadata analysis that might be implemented on the server in an attempt to detect illegal patterns of behaviour.

\subsection{Content available to server}
As we will see in due course, some types of child sexual abuse activity are most effectively identified through scanning of content.
If content is not available to servers then scanning technologies cannot be implemented there.
This property is closely linked to the implementation of features which remove the services' ability to see content.

\subsection{User authentication}\label{section:serviceuserauth}
Service providers often seek to authenticate users of their products and individuals wish to be able to verify the identity of those they are communicating with.
A variety of technical mechanisms are employed by communication services today, including private `identity keys' which are typically held only on a single client device and for which the server guarantees the association of the client with the user account; web of trust and similar designs in which users are responsible for understanding who they trust based on an external authentication channel of their own choice; and PKI (Public Key Infrastructure) and IBE (Identity Based Encryption) systems in which a delegated Certificate Authority or Key Management Server provides assurance of the identity of clients.

All of these can provide reasonable assurance of the authenticity of the peer accounts (and in some cases devices where they hold the identity) but are reliant on strong processes for human identity assurance in order to provide genuine authentication of the human users that are associated with the system accounts, rather than just asserted identities. 
These are processes that are rarely present during account creation on services today as they would create friction in the user sign up process, which is contrary to most services' business interest. 
The way in which such processes could realistically be implemented varies significantly with the type of user and/or device authentication present in the service.

While providing assurance of user attributes such as age and identity assists users in being confident that they are communicating safely, it is likely that we will have to live with asserted, unverified identity and key characteristics in the main. 
For the avoidance of doubt, the authors are not advocating for or against assured identity verification on public services and we are aware of the significant downsides of mandatory identity verification in the general case\footnote{There are specific settings, for example in the context of educational or business specific services where strong identity verification is desirable; this affects the safety properties of such services, but does not indicate the suitability of identity verification in public services where it may, for example, restrict access, aid censorship and inhibit free journalism.}.

\subsection{Message attribution}
Whilst this may not affect the risk of child sexual abuse activity to users directly, in order to act on the detection of child sexual abuse material, there may be legal and policy requirements that the material can be reliably attributed to the sender.
Though simple, in principle, for the service provider given that they are facilitating the communication, this may pose privacy concerns and some service providers offer anonymity to their users as a feature.
Some services already implement techniques such as cryptographic message franking (for example, \cite{message-franking}), which allows the service to know that a message reported to it was genuinely the one sent, without being able to decrypt the original message.

\subsection{Message longevity}
Some services are designed to provide long-lived communications that will be available to users for the duration of their use of the service.
Others implement automatic deletion of messages (sometimes termed `ephemeral messaging') -  often within seconds or minutes after the message has been read by the recipient.
In this case, there is only a limited window of opportunity for a user or client to report illegal content.

\subsection{Centralised vs distributed infrastructure}
In a typical centralised environment, servers are administrated by the service provider.
This gives it control over the routing of messages and the analytics that are run server-side.
In a distributed environment, servers may be run by any party, and network together to form the service (indeed, some services are entirely peer-to-peer with clients acting as servers).
In this case, there is less opportunity for any safety mechanisms to be implemented on the server.

\subsection{Client diversity}\label{section:clientdiversity}
Some services seek to limit access to clients over which they exercise control.
Others provide a publicly documented and accessible API allowing anybody to develop their own client.
Client-side safety mechanisms will be less universal and less effective in designs more closely resembling the latter of these scenarios than the former.
Conversely, systems that require a particular client be used necessarily offer less in the way of privacy guarantees, particularly where that client is proprietary and in the control of the same organisation as the servers through which messages are routed.
It would be entirely within the gift of that entity, were it suitably motivated, to gain access to the content without this being readily detectable to users.
This places an upper bar on the level of privacy that can be provided by such a service.

\subsection{Client Environments}\label{section:clientenvironment}
Some safety mechanisms involve running code on client devices.
The integrity protections that can be applied to that code, and the assurance that can be gained that it has been executed without interference, depend on the platform on which the client is running.
For example, browser-based clients can offer very little in the way of assurance that their code is executed as intended, but many modern mobile platforms make arbitrary modification of an installed app hard. 
There is also the issue of keeping any client up to date, in order to fix vulnerabilities and optimise performance.
Browsers always use the most up-to-date version of the client side code (since it is downloaded from the service at the start of the session) and many platforms are now reasonably good at pushing updates in apps to client devices. However,  some  methods of app installation, for example side-loading, offer little opportunity for the software to be regularly and reliably updated.

\subsection{Moderation and response}
Whether the platform in question has a reporting and moderation function has a significant effect on user safety. As important is the range of actions that can be taken as a result of such a process, from account removal to reporting to safety organisations and law enforcement, which are often policy choices made by the platform owner. 
If there are moderation services, to be effective the moderation team needs to be properly resourced and have access to the necessary tooling and data. Moderators also need to be trained in the specific content and harm types they are seeking to moderate and, in the case of child sexual abuse, they will need psychological support. 
Regulations and laws differ between countries and so moderation teams need to understand the specific nature of the child sexual abuse offences and the specific country requirements related to them.
Finally, there needs to be oversight and appeal mechanisms that are appropriate for the specific mechanism of moderation on the specific platform.

\subsection{Use by children}
Whether a service is designed for use by children or is desirable to them so that they seek access obviously has a significant effect on its utility as part of online child sexual abuse. If there are no children present on a platform, many of the harm archetypes cease to be of concern.  
In some cases, this may be equivalent to making statements about user authentication and characteristic verification (in particular age verification) as discussed in section~\ref{section:serviceuserauth}.

\subsection{Business model}
The financial drivers of the organisation running the service impact on the ways it is motivated to consider user safety.
Most widely used communication services are free at point of use and raise money either by philanthropic donation or by gathering and monetising data on their user base.
Some services that are available only on specific hardware are funded through the sale of the hardware and associated services.
Each of these business models will drive different behaviours in the service owner and also drive their view of risk of implementation, or not, of any given user safety feature. 
Even if two services are broadly similar from a technical implementation point of view, their different business models may make a particular safety feature highly attractive to one and highly unattractive to the other.

The wider incentives that a particular business model engenders, and how they align or not with the safety agenda, is important in this discussion. 
For example, some services that market themselves on anonymity of their users are unlikely to be well incentivised to modify their service to build safety systems.\footnote{We acknowledge that for some users in some circumstances, anonymity is, in and of itself, a safety feature. We do not seek to suggest that anonymity on commodity services is inherently bad, but it has an effect on the child sexual abuse problem.}
Conversely, a vertically integrated hardware (e.g. phone) and service vendor may be highly incentivised to build safety systems, in order to drive high-margin hardware sales (`The Widget Phone is the safest phone on the planet!').
When considering the overall safety implications of a given system, the underpinning business and governance models may be important factors.

\section{Current Mitigations}\label{section:current}
Having considered the range of offender behaviours, we proceed to examine the mitigations that are implemented on services today in order to prevent and detect child sexual abuse activity.
There is a large diversity in features that affect the risk of child sexual abuse activity that are implemented by different services, suggesting that a proportionate response to the child sexual abuse threat will have differing characteristics on each service.
It is not surprising, therefore, that we see a wide variety of approaches to mitigating the threat of child sexual abuse by different services, with some taking a proactive stance with a suite of features to enhance safety for their users, and others taking the view that it is a problem for their users to identify and resolve independently.
\par
We will see that the techniques described offer different levels of efficacy against different types of abuse, and that their performance is often dependent on the features offered by and overall design of the service into which they are integrated.
No single solution addresses the diverse types of harm and, where any mitigations are implemented, it is not uncommon to see a combination of these techniques applied to counter the range of threats.

\subsection{Content dependent mitigations}
Services with access to content as it traverses their servers carry out content scanning to address some of the most serious harms to users such as child sexual abuse activities of various types.
For the purposes of this discussion, we will focus our attention on how the techniques are used today to address threats related to child sexual abuse.

Hash matching and other related technologies will identify exact or, in the case of perceptual hashes, near matches to previously seen content (usually images or video) that has been classified by a trusted source (typically one or more NGOs) as illegal.
These approaches have very high precision, and matches are usually subject to human review before any reports are made to the appropriate authorities.
They are effective in the detection of previously seen content -  particularly in addressing offender-to-offender image sharing, as detailed in section~\ref{section:imagesharing}, and many instances of grooming, as detailed in section~\ref{section:grooming}.
Intrusion into the privacy of users is minimised since false positives are rare.
There is a small risk that the child safety NGO may have mis-classified an image, but the human review step mitigates the consequences of this, along with the impact of false positives from the detection algorithm.

Additionally, machine learning models may be used to classify content to identify previously unseen child sexual abuse material or conversations that are likely to be related to child sexual abuse, for example grooming as described in section~\ref{section:grooming}.
Models may be built on the images/video stream alone, the text of conversations or a combination of both which can provide additional context for the classifier.
Such techniques give services the ability to detect new child sexual abuse material as it is produced or shared for the first time and the ability to detect conversations that are likely to lead to a child sexual abuse related outcome.
In practice, these are deployed with parameters that give high precision but this type of technique will always produce significant false positives so human moderation is usually employed to prevent innocent content being reported to the authorities\footnote{This is not to suggest that human moderation in controlled circumstances is itself risk free or not a potential intrusion into a user's privacy. However, we believe mitigations for these risks are both possible and relatively simple, as we will explore later.}.

Since these techniques analyse the images themselves or the text of conversations, and can be verified by human review, they are suitable for providing intelligence that serious offences may be taking place, allowing for highly intrusive actions necessary to ensure the safety of the victims and to attempt to prevent any further offending to be undertaken.
A recent survey of service providers demonstrated that whilst offering limited benefits for other types of harms, scanning of content was considered the most effective tool for addressing child sexual abuse imagery \cite{pfefferkorn-survey}.
This survey of various service providers clearly illustrates the distinct characteristics of the detection of child sexual abuse material when compared with other harm types, with content scanning being especially useful to counter this threat.

\subsection{Behaviour dependent mitigations}
Some harmful user behaviours can be reasonably accurately detected by automated analysis of the context in which the communications are taking place without any need to examine content.
These algorithms are typically run on the servers of the provider which will use contextual information or public content (such as group names or profile images) to make determinations about whether or not to block the content.
Depending on the information available to feed these classifiers, they may be able to achieve high levels of accuracy, particularly in the case of highly prevalent behaviours (such scamming or spamming), for which there is a large, reliable set of truth data (i.e. high quality, human labelled data), and the characteristics of the offending population are amenable to such analysis.
In other cases, they may not be able to achieve sufficient accuracy which may lead to overwhelming the moderators or over-blocking innocent content.

Data analysed (depending on their availability in the context of the service) can include usernames, profile pictures, messaging behaviour (to/from/time/size), network address, use of anonymisation services, public posts, group memberships, etc.
It is also worth noting that some services implement link-shimming\footnote{Link shimming is a process whereby links are replaced (normally invisibly to the user) with a link to the service provider which, when loaded, will redirect the client to the original URL.
This was originally intended to minimise data leakage and protect against malicious content, but it also provides services with the ability to track the browsing activities of their users.}  in their client software, in which case links that are opened by users (or, indeed, automatically pre-loaded by clients) are also potentially visible by the server for inclusion in this analysis.
The output of behavioural analysis is typically tracked over time, and if it is detected to drop below a certain threshold, some follow-on action may be enacted.

Many services already perform this sort of analysis to identify offending use of their platform.
Facebook have developed robust AI techniques to identify what they term abusive accounts -- referring to spamming, scamming, fake or compromised accounts.
These techniques are reasonably effective at reducing the (high) volume of abusive accounts that they had been unable to classify by more conventional techniques, reducing their prevalence from 5.2\% to 3.8\% of all accounts on their platform \cite{facebook-dec}.
However, this approach tackles high volume `abusive' behaviours for which a large body of training data exists.
We will explore the applicability of such techniques to detection of child sexual abuse related activity in section~\ref{section:AI}.

\subsection{Identification and age assurance}
Understanding a user's identity or age allows a service to take proportionate action to ensure that the user has a reasonably safe experience.
Self-declared age is unreliable \cite{voco-report}, so some platforms employ age verification (such as checking government issued ID), or ongoing age assurance by applying machine learning models to user behaviour and user-generated content to estimate the user's actual age.
There is little objective data to understand the performance of these techniques, but anecdotally many commodity platforms claim that it is hard to reliably determine a user's likely real age.
This has a disproportionately adverse effect on utility, as it means that it is harder to justify more restrictive actions for accounts that are erroneously identified as belonging to children (as well as allowing accounts of children to access inappropriate features).
\par

The techniques described so far involve analysis of the content, metadata, or user identity to provide protections to users; we proceed to set out solutions that involve changes to the features of the service that are presented to users, offer support to users in keeping themselves safe as they use the communications service, or understand and influence how offenders may use the service.

\subsection{Restricting service features}
By limiting access to some features of the service to certain groups of users, it is possible to improve protections for vulnerable users from undesirable interactions.
Examples include restricting adult users from contacting children who are not already known to them by some metric, age-gating dating services, or restricting private messaging to users over a certain age \footnote{For example, TikTok Direct Messaging is only available to account holders who are 16 or older: \href{https://support.tiktok.com/en/using-tiktok/messaging-and-notifications/direct-message-settings}{https://support.tiktok.com/en/using-tiktok/messaging-and-notifications/direct-message-settings}}.
This can help provide some protection against offender-to-victim grooming, detailed in section~\ref{section:grooming}, by limiting opportunity to access potential victims, but it cannot be seen as a complete solution since:
\begin{enumerate}
\item{The presence of such age restrictions, as well as other social drivers, can incentivise children to enter false age information, and in-service age estimation is claimed to be hard by the platforms.
See, for example, \cite{voco-report}.}

\item{Offenders will often pose as children on the platform, building believable personas, to try to engender trust.}
\item{Grooming and abuse also takes place in the context of adults that are already known to children.
See, for example, \cite{NSPCC-grooming}.}

\item{Children often share handles and account names on other, public social media platforms and there exist `auxiliary applications' on many services, offering a `find-a-friend' service, even when that feature is not built into the platform, possibly bypassing some platform safety features.
The auxiliary applications normally require registration and may or may not work in tandem with the platform through APIs, but work unpublished at the time of writing but seen by the authors suggests many of these applications are unregulated and platforms and appstore providers do little to vet or understand the impact of these services.}
\end{enumerate}

\subsection{User reporting mechanisms}\label{subsection:user reporting}
By providing opportunities for users to report unwanted or negative behaviours, they can participate in maintaining a safe environment for all.
This is an important component of safety provision in many services but is more effective for some types of harm than others.
In the context of online child sexual abuse relying on user reporting alone is problematic for several reasons, the main one being that it puts all the onus on the victim to report abuse.
Victim-focussed charities such as the NSPCC and Marie Collins Foundation have explained that user reporting is a significant and inappropriate ask to place solely on these victims, who are likely to feel embarrassed about the situation, scared about what the abuser might do, scared that they might get into trouble with law enforcement or scared about their friends and family finding out about their situation or seeing sexual content they have created.
On top of all these barriers to user reporting of online child sexual abuse, in many grooming scenarios the victim may not even realise that they are in an abusive situation until significant harm has already occurred.
User reporting would also not cover the issue of offender to offender sharing of child sexual abuse material.
\par
User reporting services are relatively low-cost to maintain and, when well-designed, offer users clear opportunities to report harmful activities, including child sexual abuse.
It is the view of the authors that unless there is some particularly clear reason that this functionality must not be implemented by a service, it is ridiculous for a commodity service to fail to offer at least this level of support to its users, yet many fail to do so today.
An exposition of the current reporting best practice on different platforms is available in \cite{C3P-best-practice}.

\subsection{Behavioural prompts and user education}\label{subsection:behavioural-prompts}
Machine learning models run on content and/or metadata can be used to identify risky behaviour and trigger an educational message to the user.
This could provide some mitigation to offender-to-victim grooming, as detailed in section~\ref{section:grooming}, and on-demand contact abuse, as detailed in section~\ref{section:livestream}, though some users will either choose to or be convinced to ignore such warnings.
There are analogues to other `educational prompts' which have unintended consequences, such as cookie warnings and SSL certificate warnings.
`Warning fatigue' is a significant sociotechnical problem in these situations, so the design of any of these mitigations will be critical.

However, the potential deterrent effect on offenders should not be underestimated, even though this seems to be effective in only a small proportion of the population. In its annual report \cite{LFF-report}, the Lucy Faithfull Foundation describes a system that was built with MindGeek that provided a deterrence message if users searched for child sexual abuse on any of the company's global (legal) adult entertainment sites. Between early February 2021 and March 2021, nearly 22,000 users clicked through the message to the Foundation's website.
At the time of writing, more detailed statistics around engagement with the deterrent content has not been published, but is expected in due course.

More generally, educating children and parents about the risks of sharing sexual images online, and of grooming or extortion, is of critical importance to equipping them to appreciate these risks and take steps to avoid harm, both by reducing the creation of first generation child sexual abuse images and videos, and by raising awareness of the availability of effective reporting services.
However it is important to note that sadly a large number of children do not report images that they regret for a variety of reasons including shame, coercive control, lack of trust in the authorities, and disbelief that any effective action will be taken as a result.
\subsection{Understand offender behaviour}
In some situations, it is possible for certain entities\footnote{Not just entities of the State. There is some commercial activity starting in this space.} to infiltrate and understand the modi operandi of groups of malicious users, to enable their behaviour to be better understood to help drive mitigations.
Such techniques are an important component in discovering offender behaviour but do not scale easily and will not reliably identify all offenders.

\section{Approaches to Tackling child sexual abuse in encrypted Environments}\label{section:approaches}
Where end-to-end encryption is deployed on a commodity service, content is not available for analysis on the server, eliminating the most effective mitigations in use by services today to detect child sexual abuse. It is for this reason that we single out end-to-end encryption as being different in effect and impact to other service design choices.
Many of the other approaches to mitigation of child sexual abuse threats remain possible in an end-to-end encrypted environment, although often with decreased performance.
In this section we explore the options that services may wish to consider as they design the overall package of mitigations to address child sexual abuse threats to their users.
We will mirror the presentation of the mitigations set out in section~\ref{section:current}, attempting to draw out the particular implications of end-to-end encryption on the techniques.
We do not seek to endorse any of these mechanisms in particular, and wish to stress that a range of approaches will likely be necessary to provide a safe environment on any given service.

\subsection{Content dependent mitigations}\label{subsection:content-dependent-e2ee}
In an end-to-end encrypted service, content will not be available on the server for scanning, so the client will need to participate in some way for scanning of content to identify child sexual abuse.
The appropriate action to be taken following positive identification will be highly context-dependent, and could range from user notification or automatic blocking of contact, through to notifying the service, or, if appropriate under the terms and conditions of use, automatically reporting potential child sexual abuse material to a moderation team. 

\subsubsection{Scanning of text}
Analytics could be run client-side on non-media content (e.g., text or links) to establish likely intent to cause, in the case of offenders, or be at risk of, in the case of potential victims, child sexual abuse related harms.
Fairly accurate identification of grooming may be possible by this means \cite{detect-grooming}.

\subsubsection{Scanning of media}
These approaches provide detection of child sexual abuse images and videos directly. We begin by setting out some common properties before describing the range of technologies that could be considered.
\begin{itemize}
  \item{Limitations\\
 There are ways that offenders could avoid detection by any of the solutions proposed.
Whilst this limits their effectiveness in isolation, they would likely be deployed as part of a suite of measures as we will explore later. Furthermore, the intent behind their use would not be solely to catch the most committed and capable offenders, but also to detect offenders who are at risk of climbing the offender hierarchy and taking action to dissuade them from committing more serious harms.
That the techniques would not catch all offences is inevitable, and need not be seen as prohibitive to their operation. }
  \item{Dependencies\\
 Each of the techniques in this section relies to some extent on a trusted list of files that have been verified as child sexual abuse material.
The NGOs currently involved in the maintenance of such lists could continue in this role, and the precise model could be evolved to help support new techniques as necessary.
Cryptographic techniques (such as a Merkle Tree) could be employed to provide guarantees that the list had not been modified by a malicious entity.}
  \item{Practicality\\
 Many of the techniques described below are either equivalent to, or extensions of, techniques deployed in related fields today.
 However, their practicality in the context of an app running on a constrained device, integrated with a low-latency scaled messaging service, can only be proven through practical demonstration
 We could attempt to set out in detail the performance characteristics that suggest the techniques we set out below have the potential for practical implementation\footnote{We state explicitly those which are not practical with currently available technology.}, but this would require a volume of analysis amounting to a supplementary paper, which nevertheless would fail to present a compelling argument since the details of performance and scaling properties are particular to the specific implementation and its integration into a communication service. 
 On the other hand, we can point to practical demonstrators that implement techniques of several of these categories, such as those that have been developed via the UK's Safety Technology Challenge Fund.}
\end{itemize}

\subsubsection{Matching techniques}\label{section:matchingtechniques}
These techniques compare media with a set of known child sexual abuse images or videos.
Since this set cannot, for obvious reasons, be supplied to either client devices or to service providers, representations of it such as filters or lists of hashes are provided and the appropriate processing applied to content before being matched.
It is possible to use either exact, cryptographic hash matching (which provides very high precision) or robust hash (sometimes called perceptual hash) near-matching, allowing modifications of an original image to be detected with some increased probability that innocent images may also be incorrectly detected.
In order to be applicable to high-volume services and to reduce requirements for human review and consequential impacts on user privacy, when used in non-end-to-end encrypted contexts today these near-matching techniques are configured with very high precision; similar considerations would apply in an end-to-end encrypted context.

Where perceptual hashes are used (at present server-side), current designs are widely vulnerable to adversarial attacks \cite{squint-hard-enough}, although the authors know of no real-world attacks against current perceptual hash systems in over a decade of widespread use.
This does not necessarily preclude their use, but means that either new designs need to be developed\footnote{It remains an open question whether perception itself is inherently smooth, which would render the techniques employed in \cite{squint-hard-enough} universally applicable to perceptual hashes - though not necessarily always computationally practical.},
 or that mitigations to the possible attacks should be deployed alongside them.

There is a range of different options for how to implement the processing and matching of content. The following discussion is broken down according to a partitioning of that range and should be seen as illustrative of the diversity of solutions that could be considered rather than an exhaustive categorisation.
\begin{itemize}
\item{Client-side hashing with client-side matching\\
Both the hashing process and the hashed database of CSAM is contained on the client device.
Whilst we have noted that evasion by determined offenders is inevitable for any of the media scanning techniques described, putting the hash algorithm onto the client device would open it up to reverse engineering making it more likely that techniques to evade it rapidly and conveniently would be developed.
This design choice would also increase the load on the client device as it has to both perform the hashing and host the database of hashes.
In order to reduce storage requirements on client devices, it is likely that such a solution may elect to send only a (possibly random) portion of the hash database to each device, and rely on probabilities to match sufficiently well.

This approach affords maximum privacy to the user as both the image and the hash never leave the device unless a match occurs.
However, hosting the database on the client's device may expose it to unauthorised access or disclosure, with particularly impactful consequences if the hash algorithm has weaknesses allowing images to be partially reconstructed.}

\item{Client-side hashing with server-side matching\\
In this scenario the client performs the hashing algorithm and sends the hash to the service or a third party to check against the database.
This protects the sensitive database from unauthorised access.
When using a single server for matching, the image hash is revealed to the server so is no longer private.
In this case, the identity of the server's owner or administrator may make a difference in terms of perceived privacy, for example contrasting the service provider with a national organisation for preventing child sexual abuse.

It would be possible, at least in theory, for a malicious server to build a database of all known images, and thereby to gain the capability to discover which images are being shared in the majority of communications.
This would be both expensive, and would, at least in the context of a centralised service, require the service to go to great effort to undermine the end-to-end encryption it has chosen to implement, when cheaper ways of achieving this are possible (for instance through insertion of hard-to-detect vulnerabilities into client code), however the possibility of such an attack may be sufficient to undermine user confidence in the privacy protection afforded by this mechanism.

A further attack type exploits weaknesses in the perceptual hash function which may allow the server to identify features of the original content from the perceptual hash.

To mitigate against these threats, a multi-party compute approach could be used for the matching, meaning that each server would get a share of the media hash and collaborate to perform the matching process.
This increases privacy as the servers would not be able to recover the image hash without all of them colluding but comes at the cost of increased communications between the servers that would lead to increased latency for the client.}

\item{Hashing with client-server multi-party computation\\
This allows the server to collaborate with the client to compute the hash, in a manner that (depending on the specific algorithm) can reduce the amount of the hashing algorithm available to the client, and image data to the server.
The distinguishing feature of this family of techniques is that the client does not get to see the hash database, and that the server does not get to see (more than a very small amount of) information about the image\footnote{Exactly how much information depends on the scheme and the implementation details, 
but we are typically talking about a small number of bytes - not enough to be able to derive any meaningful inferences about the image itself.}.

For cryptographic hashes, techniques are available now and implemented into widely used products to provide, for example, assurance that user-selected passwords have not previously been included in password leaks \cite{google-password-check}, \cite{microsoft-password-monitor} -- these techniques are equally applicable to exact matching of hashes of images against lists of known child sexual abuse material.
There are also techniques that don't strictly involve hashing, such as the use of filters built directly on the child sexual abuse content, that can provide essentially equivalent functionality, but may also provide enhanced security of the child sexual abuse image database.

Homomorphic encryption schemes also provide a means to implement this mechanism with perceptual hashes, but further research is required to reduce the computational cost of such an approach before it is practical. At present schemes of the type to be discussed in the next section appear to offer similar (but not quite equivalent) functionality for less computational cost.}

\item{Hashing with server-server multi-party computation\\
Here the client submits the (pre-processed) media in an encrypted form, which is shared between multiple servers who collaborate to produce the hash while never revealing any image data.
The hash can then be compared with the child sexual abuse image database, either by a single server once the hash has been revealed or using another multi-party computation approach as discussed above.
In either case the image is not revealed to the server, protecting the client's privacy, and the hashing algorithm is not revealed to the client, protecting against reverse engineering.
It is likely, depending on the specific algorithm, that a multi-party computation approach will lead to increased latency. }
\end{itemize}

\subsubsection{Content Classifiers}
The techniques described above involve matching derivatives of known images. 
A complementary family of techniques involves building a model to classify media as child sexual abuse material or not based on features it contains such as nudity, age estimation based on facial or body characteristics, and body positions.
This allows detection of both known and previously unseen imagery.
Such techniques are already widely available in forensic tools and used by non-end-to-end encrypted services\footnote{Thorn's Safer and Google's Content Safety API both offer such a service, for instance.} to identify previously unseen child sexual abuse material.

Content classifiers need to move onto the client (in the same manner as matching techniques) in order to protect the privacy properties of an end-to-end encrypted system.
Whilst it is undeniable that the constrained environment in most client devices will affect the size of model and quantity of processing that can realistically be applied, with consequential impact on model performance, client-side models are already applied to detect spam, fraud and other harms, and to provide convenient categorisation of images for users. Referrals to larger, server-side models could be part of a wider solution, in some circumstances. Care must be taken to consider model inversion attacks \cite{model-inversion-image} and training data set membership inference attacks \cite{membership-inference}.

It is possible that the models distributed to clients could be used by offenders to scrape the web for child sexual abuse images, however a similar argument suggests that the more effective and larger models used on servers could be applied by NGOs or Law Enforcement to identify and enforce removal of this content\footnote{In the same vein as existing approaches based on hash matching, such as Project Arachnid: \href{url}{https://projectarachnid.ca}.}.

Machine Learning classifiers could also theoretically be used in conjunction with client-server multi-party computation to provide privacy protection to both user data and server classifier data. However, the performance characteristics of these techniques are too expensive for immediate deployment \cite[Table~6]{xnorr}.

\subsubsection{Safeguarding content}
Services could, at the request of a child or their parent/guardian or by default for all children, add safeguarding functionality to a child's account, giving some other party access to content (and/or metadata), facilitating supervised use of the service.
We note that this may drive some children away from services that implement these sorts of techniques.

To provide this type of access to content, services could add an additional user to each end-to-end encrypted conversation, allowing for server-side supervision of vulnerable users. It would be technically possible for the safeguarding user to be another user rather than the service or a responsible third party.  We note that this possibility is often undesirable due to the implications it could have for children with abusive or controlling home environments and the fact that it could provide a false sense of security given that many parents are incapable of providing such moderation services. It may also create socio-technical vulnerabilities that could be exploited by offenders. 
Such action could be made explicitly visible to all parties involved, which may provide partial mitigation against the capability being misused against other users.

Whilst only effective when taken up by the user community, it could provide strong safety protections - potentially equivalent to those in non-end-to-end encrypted environments - for children to whose accounts it was applied. 
It is plausible that hackers could seek to exploit the interface that controlled the status of safeguarding user on accounts they wished to access. However, to gain benefit from this in the model where the service provides moderation, they would need ongoing access to servers, which in most cases would constitute a higher complexity of attack than seeking to attack the client devices from which they wished to gain content.

It is also worth noting that this kind of mechanism may place some children at additional risk from abusive or manipulative parents, even when the parents themselves don't have access to content, and whilst the technique would be technically relatively straightforward to scale, research would be necessary to determine how well it would be likely to cover the users most at-risk and how at-risk children could be effectively protected.

\subsection{Behaviour dependent solutions}
\subsubsection{Automated metadata analysis}
As mentioned above, some services perform ongoing server-side classification of user metadata or public activity.

This is still possible regardless of the content encryption implemented by the service, but may be rendered less effective, for example if information such as file types or sender details are not available to the server. 
Classification such as this may have uses as a course technique for identifying where more intrusive techniques could be targeted, but will not address the child sexual abuse threat alone as it will always be significantly limited by the ground truth data it can access which will determine whether the actual behaviour associated with the metadata is benign or harmful. 

\subsubsection{Safeguarding metadata}
Alternatively, providing parents/guardians with a view of metadata relating to their child's use of a service may allow them to manually assist in safe use of the service, for example in accurately identifying their child's peers.
Such mechanisms typically wouldn't require change to existing end-to-end encrypted designs, except when sender identity is concealed from the server.
This would only be as effective as the parents'/guardians' ability to spot anomalous behaviour, and offenders may be able to circumvent it by presenting themselves as acceptable peers.
Sadly, as well as those who are not technically proficient enough to act in this capacity, there is also a subset of parents/carers who are not sufficiently motivated to proactively act to ensure the safety of their children online (including some who are themselves offenders, or who may cause other forms of harm to their children).

\subsection{Identification and age assurance}
Ongoing age assurance on commodity platforms can be hard, with factors such as the child's incentive to lie (in order to get access to restricted content or services) being important considerations. 
This is likely to be harder without access to content and this is certainly true on platforms without public areas that can be mined for data that implies age.
 However, as noted in \cite{voco-report}, content is but one input into an age assurance system and it is unclear whether sufficiently accurate age estimates can be made without it. 
What is more certain is that further research is needed in this area, and that platforms would need to be incentivised to implement such systems if they are considered to be desirable. 

\subsection{Restricting service features}
Limiting access to risky features to either children or those who are identified as potential threats could help mitigate some harms. While not the primary use-case in mind, it would certainly be feasible for end-to-end encryption to not be applied to conversations between a child and new contact, for example.  
Techniques of this nature apply regardless of whether a service is end-to-end encrypted or not, but may be particularly important in the case of end-to-end encrypted services as when users are informed of the presence of encryption, it is likely to result in increased perception of security and trust \cite{e2ee-perceptions}, potentially leading to riskier behaviour.

\subsection{User reporting mechanisms}
Whilst it cannot be relied upon as the only solution for child sexual abuse harms (potentially unlike other harm types), user reporting and moderation is an important part of user safety systems. 
In the context of an end-to-end encrypted service, it is particularly important that user reporting functions are clear that the content of the conversation will be revealed to a moderator so that users can give informed consent.
With that in mind, however, there is much that could be done to make user reporting functions more accessible and reducing barriers to their use, such as:
\begin{itemize}
  \item{Reporting built-in to client applications, with options to directly upload the previous $n$ messages for human moderation.}
  \item{More neutral questions.}
  \item{Prompts to users to consider reporting potentially unsafe behaviour that the application has detected.\\
    Such prompts may be triggered by any of the other approaches outlined in this section, and may vary in their implementation depending on the circumstances involved.
    For instance, a conversation identified as likely to present a significant risk of grooming for sexual abuse could be flagged to the potential victim (but not the potential abuser) as such, offering them an opportunity to report the conversation for moderation.}
\end{itemize}
As highlighted in section~\ref{subsection:behavioural-prompts}, where users are presented with warnings or prompts for action, these must be reasonably reliable and timely so as to avoid warning fatigue.

Regretably, even where reporting functions are readily available today, many children who have been subjected to sexual abuse via communications services do not report it, and in some cases may not even be conscious that it has occurred until significantly after the event, or until they after are under the coercive control of an offender.  

Whilst in most cases it is anticipated that a user would be involved in a reporting decision, there may be situations in which it is more appropriate for a client to choose to report activity for moderation without requesting specific approval from the user to do so, for example following highly-reliable hash matching, or where a child is identified as being in imminent danger with high confidence.
Situations in which this could be considered would include the detection of sharing of known child sexual abuse content between offenders, known abuse content received by a child in the course of grooming for abuse or a grooming offender making immediate plans to meet a child for the purposes of contact abuse.
Care would need to be taken in such cases to balance the risk of the moderation team being swamped with innocent content, with consequent loss of privacy for those accounts that were referred.

\subsection{Behavioural prompts and user education}
As in section~\ref{subsection:behavioural-prompts}, situation-dependent prompts may be offered to users to help them to identify potentially harmful situations.
The difference in an end-to-end encrypted context is that any analysis on content required to provide accurate assessment of the risk of the situation must be performed in a manner that maintains the privacy of the user content from the server, likely using one of the techniques outlined in \ref{subsection:content-dependent-e2ee}.

\subsubsection{Referral}
Directly referring potential victims to support services or charities may enable them to seek help that they otherwise might not have obtained.
This can be achieved without any compromise to the users' privacy, but accuracy when such actions are taken is still important as otherwise users may perceive that legitimate actions are being discouraged.
This referral could be to platform moderators and to charity helplines such as those run by the NSPCC and the Lucy Faithfull Foundation. Lack of friction in this referral process will be critical to its utility. 

As shown by the recent work of the Lucy Faithfull Foundation described in section~\ref{subsection:behavioural-prompts} there may be utility in providing low-friction referral services for offenders as well. While we are under no illusion as the overall efficacy of this intervention, it remains a useful, if limited, tool. 

\subsection{Understanding offender behaviour}
End-to-end encryption is unlikely to have a major effect on the mechanisms used to understand offender tradecraft in use today, since they rarely rely on covert access to private content of targeted user groups.

\section{Evaluation of client-side image scanning}
Whilst client-side content scanning is only one of the range of potential technological approaches we have outlined in section~\ref{section:approaches}, it is nevertheless the one which has been the subject of particular attention in recent analyses, especially following the announcement and subsequent withdrawal of Apple's proposed implementation for iCloud \cite{neuralhash}.
Many publications and commentators have taken a binary approach to analysis of client-side image scanning techniques, looking at them in isolation, rather than part of a wider sociotechnical safety system \cite{biop} \cite{cdt-outside}.
There is also an unhelpful tendency to consider `end-to-end encypted services' as academic cryptosystems, rather than the set of real-world compromises they really are. 

Commentators have referred to the privacy promises of end-to-end encryption ~\cite{cdt-outside},~\cite{bsr-human-rights}, suggesting that any end-to-end encrypted system provides a guarantee that only the sender and intended recipients may access content unless one party takes specific action.
We do not believe this academic definition is appropriate in the context of user safety on scaled commodity services.
Firstly the business model of many advertising-driven services seeks to break user privacy at scale and removing access to content from these services does little for user privacy.
Secondly, the `social contract' between users of a service and the service owner, even in an end-to-end encrypted world, is not merely about access to content. A survey by ECPAT of 9,410 adults across 8 European countries in September 2021 shows that over 75\% of those surveyed believed that detection of child sexual abuse on a platform was as important or more important than personal privacy \cite{ecpat-beacon}.

In our view, a meaningful assessment of the risks and benefits of using client-side scanning can only be performed in the context of the service into which it is integrated; indeed it is more significant to discuss the risks and benefits of the system as a whole, rather than of one component of it.
However, there are some generic attacks against client-side scanning in the context of end-to-end encrypted service to which there are mitigations that we describe here, and some properties of client-side scanning to which we present alternative perspectives. 

Importantly, there are also potential attacks that exist regardless of whether access to the images occurs on the server, as today, or on the client, as is proposed in end-to-end encrypted services. While these have been tolerated for many years, the move to end-to-end encryption - and the removal of default access to content by the service infrastructure - seems to change the user expectation to the point where mitigations become necessary. 
We will describe the attacks, the potential impacts they may have if executed and suggest mitigations. 

\subsection{Common attacks against image scanning}
\begin{itemize}
\item{Manipulation of image database\\
With any system that tries to match user content against `contraband' content, there is a risk of a malfeasant actor covertly adding content to or removing content from the database content in order to achieve some other goal.  

The most obvious way that a database may be modified is where the curators of the database are coerced by a third party. The most common actor hypothesised here is a government, using legislative mechanisms or just pressure on individuals to gain assistance. We are reminded that there are governments around the world who would seek to use extra-judicial mechanisms to influence such a database, using it for things other than countering child sexual abuse. While we know of no law in an oppressive regime that acts extraterritorially in the manner necessary\footnote{Such that, for example, NCMEC or the IWF could be compelled to add an inappropriate image to a child sexual abuse image database at the request of the oppressive regime.}, we nevertheless accept this as a valid attack. 
 
It is sensible to assume that at least one person involved in the management of the child sexual abuse image databases actually has a sexual interest in children. We use the analogue of national security clearance here to demonstrate that no vetting process will identify all malfeasants in all cases. Therefore, we must consider insider attacks with motivated insiders. 

Finally, we consider regulatory capture. This is the phenomenon where a regulator acts in the interests of a small number of those it seeks to regulate at the expense of a much larger population. Many of the child protection charities that manage these databases are at least partly funded by the big tech companies whose services are the subject of this discussion and so it is reasonable to ask how we can be certain that the service owners are not manipulating the curation of the database. We do not believe that the incentives align to make this likely and we have not seen any evidence of this over the years. 
A more likely situation arises when we consider `mission creep' of a regulator that may be responsible for issues wider than just child safety. Then the question becomes one of how we can be sure that the database is not being used (legally but covertly) to discover other classes of image, for example terrorism images. 
}

\item{Image collision attacks\\
Here, an actor creates an image that is not child sexual abuse, but is matched as such by whatever detection algorithm is in use. This sets in motion whatever process is in place to deal with the sending of child sexual abuse imagery. All detection mechanisms have a non-zero false positive rate and so any subsequent processes are already designed with that fact in mind. Image collision attacks appear to be equivalent to the generic false positive problem.}

\item{Image seeding attacks\\
Here, an attacker seeds a real child sexual abuse image to a user or community they wish to affect.  
}
\end{itemize}

\subsection{Attacks against client-side image scanning}
Moving the image scanning and detection functionality from the server to billions of clients certainly admits new attacks and changes the risks.
\begin{itemize}
\item{Oracle attacks\\
In many practical client-side scanning systems, a positive detection event leads to the content being sent to a third party (usually a moderator). Given this must be performed by the client which is in the physical possession of the adversary, it is possible\footnote{Though not necessarily trivial - at least in theory all that need be transferred is an additional moderator key for the image and a flag to route the image to the moderation service; this may be indistinguishable from the externals of the communication with the server, thus requiring instrumentation and detailed understanding of the application to determine whether an image has been referred.} for the adversary to infer whether a particular image is part of the contraband database. This is obviously equivalent to a fully local classifier being exercised, which is the easier version of this attack. This is obviously different to the server-side scanning where it is much harder to observe the output of the detection algorithm.}

\item{Obtaining information about child sexual abuse content\\
It is possible that some meaningful information from the original content may be recoverable from a set of perceptual hashes or from a model built on child sexual abuse training data, for example through model inversion attacks \cite{model-inversion-image}. Again, this seems to be much harder in the server-side case where access to the model or a set of perceptual hashes is not intended. 
}

\item{Fragility of detection mechanism\\
The client must perform two actions in order to send an image - it must submit the image to the detection scheme and, separately, must encrypt the image using the recipients' keys and send this to the service for delivery. In the general case (since we cannot assume anything about the client in terms of execution security), it is trivial to modify the client such that the image used to participate in the detection mechanism is always benign and the real child sexual abuse image is encrypted and sent separately. This issue does not exist in the server-side case where the client performs a single action that causes both detection and routing to occur atomically. }

\item{Increase of client security vulnerabilities\\
Client-side scanning will require relatively complex additional code to be added to the client in some way. Additionally, this code will potentially be processing untrusted content - that is, images sent by a third party. If this code is not written in a defensive manner and regularly maintained and updated, there is a likelihood that remotely exploitable vulnerabilites may be introduced into the client.}
\end{itemize}

\subsection{Potential impact of attacks}
The attacks listed above can be used individually or in combination to achieve certain outcomes for an adversary. For the sake of brevity, we describe only the broad classes of attack here. We accept that there are subtleties and variations on these themes that will require analysis and mitigation in the context of specific implementations. 
\subsubsection{Generic impacts}
There exist a number of malicious outcomes that can be achieved (at least in theory) through the abuse of any image scanning system, regardless of where detection occurs. For clarity, these risks exist today with server-side image scanning. 

\begin{itemize}
\item{Unauthorized surveillance\\
Through manipulation of the detection mechanism, it may be possible for an adversary to retrieve arbitrary image content that is sent on the service. For example, an adversary could attempt to retrieve all images of cats sent by users of the service through manipulation of the contraband image database or training set used to train client-side classifiers. This would result in images of cats also being sent for moderation. }

\item{Unauthorized tracking\\
Through manipulation of the detection mechanism, it may be possible for an adversary to infer which users send and recieve arbitrary types of image content. For example, an adversary could attempt to infer all users who send images of cats through manipulation of the contraband image database or training set used to train client-side classifiers. This is similar to the previous impact, but the adversary only observes the binary decision output of the detection mechanism.  
}

\item{Disrupting free speech\\
Any moderation system can be used to disrupt free speech if insufficient checks and balances are in place. In this case, the adversary would subvert the detection mechanism such that detection of the content they wish to stop would lead to neither a referral to moderation or a moderation decision, effectively leaving the content in limbo and unsent. 
}

\item{Denial of service\\
Again, this a feature of any human-in-the-loop moderation system. Content that is detected as child sexual abuse images is sent at scale by an adversary in order to overwhelm the human moderators. The specific impacts would depend on the details of the moderation system and the service. 
}
\end{itemize}

\subsubsection{Specific impacts of client-side scanning}
\begin{itemize}
\item{Subversion of detection\\
As detailed above, there are a number of ways that an offender can modify their client to effectively subvert the client-local part of the detection system. Some will claim that this makes the whole effort nugatory. However, this is only an issue in one of our harm archetypes as we will see later.}

\item{Compromise of client\\
Since client-side scanning requires more code on the client, there is a potential issue of security vulnerabilities leading to compromise of the client and the subsequent unauthorised access to data. Implementation details will be critical to assessing this risk but, as we shall see shortly, we do not believe that this is practically an increased risk.}
\end{itemize}

\subsection{Potential mitigations}
\begin{itemize}
\item{Defence-in-depth moderation\\
In much of the published literature, there seems to be an expectation that any technical detections are automatically and directly referred to law enforcement. This is untrue today and would be untrue when client-side scanning was in place. 
The moderation and reporting pipeline in place today is designed around two axioms - that any automated detection system will have a non-zero false positive rate, and that for some content, illegality is an expert judgement\footnote{Obviously, there is also content where any right-minded person would consider it illegal.}.
The moderation and reporting pipeline today usually has three steps after automated detection. Firstly, a moderator employed or engaged by the service examines the suspect content to see if it is likely to be illegal (although a small number of service providers strangely omit this step). If it is, the information is packaged up and sent to (for US services) NCMEC, as detailed in section~\ref{section:scaleofproblem}. NCMEC then judges the report and, assuming they concur, this is forwarded to the relevant national law enforcement agency. The national law enforcement agency then decides what action to take, per their normal process. 

In this case, we shall assume that NCMEC is also the curator of the database of images (or, equivalently, the training set used to train a classifier). Even in the worst case scenario of NCMEC as an organisation being the adversary (who needs to both modify the database and observe matches) there is an independent check by the service-owned moderation process. 
}

\item{Integrity of database\\
There remains a case for ensuring that any database used for matching known child sexual abuse imagery contains only child sexual abuse imagery. Assume that for the sake of this discussion the global child protection charities that manage the various collections of child sexual abuse images can agree a single list of images\footnote{This is not the case today, but should be reasonably simple to engineer.}. In this case, it should be technically simple to :
\begin{enumerate}
\item{Publish information that embodies the state of each copy of the database. For example, constructing a Merkle tree from the images in the databases and publishing the root hash would provide assurance that each database has a consistent set of images.}
\item{Cryptographically bind any queries to that database to the published state, such that if the database is not in that state, the query will fail.\footnote{There is a requirement to put some level of trust in the charities involved in this work.}}
\item{Allow for third party audit of the contents of the database in order to show the published state represents only child sexual abuse images. There are currently legal barriers to this, but we believe they can be overcome relatively easily.}
\end{enumerate}

There are certainly other ways of assuring the integrity of the database such that it is possible to assure the public that only child sexual abuse images are present in the database. This paper does not seek to define the optimal mechanism, but only to show a proof of principle that a solution to this issue should be possible. 
}

\item{Client software integrity\\
If significant processing that is critical to safety systems is implemented on the client, then there is an immediate incentive for offenders (in some cases) to seek to undermine that processing and neuter the safety system for their device. As described elsewhere in this paper, this is not always catastrophic, for example when considering the harm archetypes where actors other than offenders are involved (for example the grooming archetype). However, details will matter about the level of work required to perform this and whether the service can detect modified clients through some sort of attestation scheme. This will, again, depend on the details of how the service clients are implemented, as described in section~\ref{section:clientdiversity} and section~\ref{section:clientenvironment}. It may be that a thick app running on a modern, high end smartphone is difficult to modify in a way that is not easy for the service to detect. It is certainly true that a browser only solution will be trivial to modify. Likely offender responses, for example using low-end devices where software attestation is not as robust, are detectable by the service. While every effort should be made to ensure the integrity of client software that implements safety features, we do not believe that the possibility of modification by offenders removes the utility of this approach.  
}

\item{Development and integration\\
Comments have been made that a client-side scanning system must introduce vulnerability because it cannot be kept up to date and patched. We do not accept this line of argument. We hope this paper has shown once again that {\em details matter} and it should be clear that safety systems will be particular to a given service. Thus, it is most likely that these safety systems will be implemented {\em by the service owner} in their app, SDK or browser-based access. In that case, the software is of the same standard as the provider's app code, managed by the same teams with the same security input. The update cycle is managed entirely by the service owner, but would be expected to be in line with their standard processes on which their users rely already. In short, we do not believe that there is anything exceptional about client-side scanning code that would mean that it would be managed differently to any other software provided by that developer. To be clear, claims that government-written code (either in source or binary form) would be embedded in commodity products are facile. No technically competent, responsible government would require or approve of that approach. 

A similar concern is that there is the potential for third parties (governments, etc.) to interfere with the client-side scanning system's software. We have considered potential mitigations to interference with training/hash data above. Whilst it is always possible that a third party could seek to leverage influence over a developer of any software, it is worth considering that client-side scanning software will be distributed to commodity devices on which providers of other software (including operating systems and antivirus) already have more powerful access\footnote{In many cases antivirus has the ability to run arbitrary code on a specific user device of the provider's choice.} than that which would be available to a client application. That the threat of interference is understood and mitigated in these more risky contexts demonstrates that it can be in the case of communications services' software, including any client-side scanning that may be implemented.
}
\end{itemize}

\subsection{Other concerns}
Other, sociotechnical issues have also been raised around the use of client-side scanning. These risks do not have simple technical solutions (to our knowledge) and deserve democratic, informed debate.
However, we do not believe that these risks are unprecedented, nor that they make it axiomatically unwise to implement client-side scanning in an appropriate context. 
\begin{itemize}
\item{Client-side scanning weakens privacy\\
This concern is central to the potential adoption of client-side scanning, and other safety technologies. We do not believe that any evaluation of privacy impacts can be performed in isolation, since the context of the service into which the safety technology is embedded will significantly affect those impacts. Open and honest analysis of a variety of client-side scanning technologies needs to be performed on a variety of candidate services to fully appreciate this issue. However, our work suggests that this risk is likely manageable in the majority of cases. 
}

\item{Client-side scanning changes a user's relationship with their device\\
There are a number of slight variants on this concern, but they seem to coalesce around the concern that a user's device may act for the benefit of a third party and potentially incriminate the user.
We believe that these legitimate concerns need to be balanced against the benefits to all users that may result from well-implemented client-side scanning, and considered in the context of the wider technological landscape into which client-side scanning would be deployed.

In all the configurations we have considered, client-side scanning for child sexual abuse images is, at least in part, for the benefit of the user: all users benefit from protection from the receipt of unsolicited child sexual abuse content; would-be victims may be warned in advance of sharing illegal images of themselves; harm to victims of abuse is reduced as the sharing of their image is restricted.
It may even be considered that offenders benefit in having their offending detected early -- allowing for preventative programmes which reduce escalation in offending and more serious consequences for themselves and for their potential victims.

It is also the case that reporting to moderators, with follow-on involvement with law enforcement where offending is believed to have occurred, is only one potential outcome of detection of (suspected) child sexual abuse images.
As we have noted, a range of actions are available for service providers to consider, taking into account the risk environment of their service.

Furthermore, analysis of user data for the detection of criminal or suspicious activity, on the user's device, is not unprecedented. Fraud detection tools and anti-virus are two examples of technologies that are widely deployed which, in at least some cases, incriminate the user of the device on which they are running through inspection of the user's data.
}

\item{There is no compelling case for the necessity of client-side scanning\\
We believe this to be false and borne from a lack of accessible information. We hope that this paper goes some way to remediating that information asymmetry. 
}

\end{itemize}

\subsection{Conclusion}
There will be legitimate concerns around the precise details of any client-side scanning system in the context of a given service.  In many cases this will require careful analysis of each implementation, just as for any other security-relevant feature of a product or service. There is undoubtedly work to be done to examine and understand the real-world effect of client-side scanning mechanisms in the context of real systems and real harms.
This paper does not intend to do this analysis, but instead seeks to motivate a more inclusive debate.
However, in our work so far, there seems to be nothing exceptional about using the client as part of a safety system for countering child sexual abuse that would render it axiomatically impossible or unwise. In particular, we believe there are reasonably simple mitigations for the legitimate concerns raised around client-side scanning. 

\section{Why AI and Machine Learning are not the answer}\label{section:AI}

Many commentators have suggested that artificial intelligence and machine learning techniques could be used to mitigate any potential loss of safety measures in end-to-end encrypted commodity services. 
The major platforms that are free (from a monetary perspective) at the point of use have a business model driven by monetisation of metadata around users, for example through advertising. 
This seems to be able to discriminate particular characteristics about groups of users in the larger population to target adverts or content that are likely to generate the `right' reaction. 
It is therefore intuitively comfortable to suggest that similar techniques could identify characteristics related to child sexual abuse behaviours and characteristics. 
We do not believe this to be true for the general case and will explore the constraints and limitations of such techniques in classification of these behaviours. 

\subsection{A concrete example}
There is little in the public domain about how major commodity platforms use machine learning and artificial intelligence to combat various threats on their platforms.
This is especially true when considering more advanced artifical intelligence and machine learning techniques that operate on the wide range of metadata available to platforms. 
Facebook has, however, published a corpus of work on their use of artificial intelligence and machine learning techniques to make their platforms safer, specifically to deal with `abusive accounts'.
Facebook's work to remove abusive accounts - those which are set up to perform large-scale scamming or spamming, for example - from their platform using machine learning is described in the \href{https://research.fb.com/wp-content/uploads/2020/12/Deep-Entity-Classification-Abusive-Account-Detection-for-Online-Social-Networks.pdf}{paper} by Xu et al `Deep Entity Classification: Abusive Account Detection for Online Social Networks' \cite{facebook-dec} and its references and these form a useful reference to the state of the art. 

It is worth being explicit that we are not singling out Facebook's work in order to suggest they are `bad' or not doing enough in relation to child sex abuse.
This collection of papers is a useful, real-world example of the application of machine learning and artificial intelligence to detection of abuse on the type of platforms that we are concerned with that also uses large-scale metadata.
It is therefore a useful concrete example that demonstrates state of the art and that is the only reason that we reference it. 

We do not suggest that the Facebook system we reference is the only way to use AI for user safety problems.
Here we intend only to show the difference between two harm classes - abusive accounts and child sexual abuse. 
The specifics of the system referenced are less important than the characteristics of the child sexual abuse harm archetypes and how they are represented in metadata, which would apply to all current machine learning and AI systems.

\subsubsection{Intent}
The system as described intends to detect `hard to find abusive accounts' where `hard to find' implies the accounts have already passed rules-based detection.
`Abusive account' in the referenced papers implies an account that is or intends to engage in one of the following harms: scam, spam, fake account or compromised account.
The intent of the system is to classify accounts as abusive before they actually engage in any abusive behaviour and either remove them from the platform entirely, or remove their ability to engage in the harmful behaviour until they act differently (e.g.  remove particular platform functions).

\subsubsection{Overview of system}
We summarise the system as described to provide background and context only.
Those interested in detailed descriptions, performance and analysis should refer to the original paper and its references.

\begin{enumerate}
\item{For every node in the social graph on Facebook, a set of features are extracted.\\
This is of the order of a few tens of thousands of features that seem to be intended to describe how the node embeds in the larger graph, how it behaves in relation to its first and second order neighbours and some categorizations of the node (possible interests etc).
We will call this very high dimensionality feature vector the `initial feature vector'.}

\item{A semi-supervised learning model is run on these features, using weakly labelled data at scale.\\
The system does sensible things to remove features that will dominate and bias the models, and the labelling is performed automatically using a number of low-precision mechanisms.
The design of the system is a set of hidden layers of neural networks, working on these `approximately labelled' training data, each linked to specific abuse modalities.
The output of the hidden layers is a feature vector of much reduced dimensionality (compared to the initial feature vector) that better represents classification-relevant features for the learned abuse types.}

\item{A supervised learning model is run on this reduced dimensionality vector, using high quality, human labelled data.\\
For each abuse type, a set of high quality, human labelled training data is used to train a gradient boost decision tree classifier (GBDT) on the reduced dimensionality feature vector output from the hidden layer neural network.
Each GBDT is now able to output a Boolean value to indicate whether a particular node in the graph is classified according to its abuse type (i.e.  the `spammer' GBDT will output a 1 if the node appears to be a spammer, a 0 otherwise; the `fake account' GBDT will output a 1 if the node appears to be a fake account, a 0 otherwise; a node will have a 0 or 1 for each of the specific abuse type GBDTs).}
\end{enumerate}
 
Once trained, the system will produce a Boolean vector for each node in the graph, representing the different abuse types that the account has been classified as participating in.

Of note is the characteristics of the training data used and the observability of characteristics that can lead to specific labels and the need for abuse-specific, high quality, human labelled training data. 
The impact of the abuse we are trying to classify must be observable in some way in order to allow human classification, which will obviously be driven by observing the abusive effect itself. 
As we will see, this is challenging for child sexual abuse harms. 

\subsection{Challenges in application to the child sexual abuse problem}
We believe there are a number of significant challenges to applying this sort of approach to child sexual abuse related harms to replace current content moderation approaches.
We are using the deep entity classification for abusive account detection on Facebook as a real example (since it is one of the few that are detailed publicly), but believe the concerns that follow will apply to any machine learning like system that operates only on service metadata.

\subsubsection{Types of harm}
All the referenced research deals with harm types that require scale, through connectivity of the node in the graph and activity of its first and second order neighbours.
The four abusive account types talked about in the paper all exhibit these properties: spamming and scamming are all about getting to as many people as possible; fake accounts are likely to want to be highly connected and active to achieve their aim (for example disinformation) and compromised accounts will normally be highly connected and active (otherwise, what is the point of compromising it?).
The paper \cite{facebook-dec} notes this limitation and suggests that harm is proportional to connectivity and activity in some way.
To quote : `However even if such accounts [accounts that have low connectivity or activity] are abusive, they inherently have less impact on Facebook and its users.'.
This is obviously not true of the child sexual abuse problem set \footnote{`High' and `low' in this case are obviously relative and different for the two harm types. While an account in a child sexual abuse case may have a hundred connections, a useful spammer account will likely have many thousands.}.
While there is probably some analogue in the case where there is genuinely non-malicious viral sharing of a child sexual abuse image, none of the other child sexual abuse harm archetypes seem to fit this model.

\subsubsection{Performance}
The claimed performance of the DEC system is impressive in machine learning terms.
The system exhibits an AUC of 0.90 which is excellent (although limited in utility as a metric in this highly biased population context).
The recall at precision 0.95 is 0.50, which is reasonable in general and useful in the context the system is used.
We do not believe that similar performance can be achieved on the child sexual abuse harm types, but for the sake of argument,  assume it can.
{\bf These best-case performance measures seem inappropriate for the harm caused by child sexual abuse cases.}
The reality in this case would be that the system would be 95\% sure accounts flagged as abusive really were abusive, but it would only be detecting 50\% of the abuse on the platform.
We expect that, in an operational context, no scaled platform would run a classifier at a precision of 0.95 due to the excessive error and so we would expect an operational precision around 0.99 with an attendant loss of recall.
As in most machine learning problems at scale, precision is critical for real world use.

It is useful to compare these performance characteristics with those of PhotoDNA, one of the perceptual hashes used to detect known child sexual abuse imagery. 
The authors are aware of large-scale, private testing of the PhotoDNA algorithm which suggests that, in a non-adversarial model, PhotoDNA has a false positive rate of around 1 in 50 billion. 
PhotoDNA was designed to be robust in the face of standard image manipulations and transformations. Under these conditions, its true positive rate is greater than 99\%.

\subsubsection{Base rate effect on performance}
The description in the paper suggests a base rate of `abusive accounts' of the order of 0.05 (i.e. approximately 5\% of the accounts, or several hundred million, are abusive at any given time prior to mitigation).
Even this rate requires resampling of training data to better model the overall population.
The base rate for child sexual abuse offender related accounts is likely to be orders of magnitude smaller\footnote{Again, we note the viral image sharing archetype must be treated separately.}.
Broadly, for a population with a low base rate, a classifier will need to have better true positive rate and false positive rate to one applied to a population with higher base rate in order to achieve comparable precision.
However, the very low base rate for child sexual abuse related accounts, coupled with the random sampling construction of the training set will lead to true positives being over-represented, causing discrimination problems.
{\bf The extremely low base rate of child sexual abuse offender related nodes will lead to excessive noise, leading to worse classifier (and overall) performance on this population than the more prevalent abusive account types.}

\subsubsection{Availability and utility of weak training data}
This section refers to the training data used during the semi-supervised learning stage on the high dimensionality initial feature vector.
The model as described requires a significant volume of weakly-labelled training data for the semi-supervised first stage model that reduces the dimensionality of the initial feature vector to a lower dimensionality vector for the GBDT classifiers.
In the main paper and its key references (\cite{noisy-learning} for example, at reference 52 in the Xu paper) these labels are generated from heuristics, rules and user generated content (including an untrusted machine classifier).
In the case of the four abusive account types considered in the paper, these sources can be assumed to be doing a relatively good job of labelling `bad' accounts.
In \cite{noisy-learning}, for example, the weak labelling already makes a significant contribution to the final performance.
This is due to the specific characteristics of the harm - the abusive accounts will all have similar characteristics (e.g. connectivity scale), the harm is likely to lead to user reporting, there are rules and heuristics that can reliably detect a reasonable proportion of the accounts in a deterministic way and so on.
In the child sexual abuse case, with a very low base rate, and fundamentally different harm types, {\bf we do not believe that the machine generated approximate labels will be useful over time.} 

It is also worthy of note that we have seen offender behaviour evolve over time in significant ways. Sometimes this is in response to new platform features or safety measures; sometimes it is due to one of the offenders being arrested. This new behaviour is often invisible to the platform until some signal is generated to alert the platform or law enforcement to the new behaviour. This could be a user report, the sending of a known child abuse image as part of this new behaviour, the arrest of an offender who then discloses new tradecraft and so on. What is almost certainly true is that these new behaviours will not be simply derived from old behaviours and so the weak, automated labelling is likely to fail in the face of offender tradecraft evolution. 

\subsubsection{Availability of high quality (human moderated) training data}
This section refers to the training data used during supervised learning on the reduced dimension output of the hidden layer system.
This system requires high quality labelled data for the second stage supervised learning.
The argument has been made that {\em without encrypted content} there will be sufficient signal to allow a human moderator to determine whether a particular event on the platform is indeed a child sexual abuse harm.
For almost all child sexual abuse harms, we do not believe that these features provide sufficient discriminatory power on non-content for the reasons described elsewhere in this paper.
This seems to lead us to the case where all high quality labelling is being generated by user-initiated reporting which, as has been discussed at length, is ineffective at scale.
Victim generated reporting in the child sexual abuse harm types will almost certainly lead to highly biased data. Reports will only be created by those who are emotionally mature enough to see the value and simultaneously not in the thrall of an offender. Reports will vary in utility depending on how far down the grooming journey (for example) an offender-victim pair have progressed before the victim realises. This will lead to inherent bias in the reporting data which will lead to bias in the labelling and the generated model. This will severely limit the types of harm that can be detected (ignoring more fundamental issues of performance noted elsewhere).
Regardless, reliance on user reporting puts the burden of responsibility on the most vulnerable and harmed.
No user interface design changes that fact.

From metadata alone, viral sharing of child sexual abuse images will likely appear to be similar to other `controversial' content sharing.
While the intent would be to `damp down' the viral spread of child sexual abuse imagery\footnote{Except, obviously, when this sharing is in public where existing content scanning tools will continue to be effective.}, `controversial' content appears to be used to promote engagement.
We cannot see how this dichotomy would be solved in the long term.
It is also worth noting that, in respect of the offender-to-offender indecent image sharing archetype, law enforcement experience shows that offenders will invariably share previously known content, which may include images that are also shared virally, in order to achieve other aims.

As offender behaviour evolves and new platform features are added, it is highly unlikely that sufficient human labelled data could be generated to make these models sufficiently accurate in practice, since child sexual abuse harms are almost uniquely content driven, and that content will not be available to generate the high quality labelled data required.
This seems to be especially problematic when new platform features are introduced where there is no {\em a priori} knowledge on which to build.
Therefore, we believe there will be {\bf insufficient high quality training data to maintain classifiers that are of use over time, in the face of adversary behaviour evolution and platform feature changes}.

\subsubsection{Adversarial model}
The adversarial model considered here is about changing node behaviour to make it look `normal enough' across all the deep features that go into the initial feature vector (or to poison the first stage training set, but the argument is probably similar).
Given many of these deep features are not in the direct control of the adversary, the argument is that it is hard to create a node that is bad, but looks good to the classifier.
This is probably true for the examples listed in the paper.
However, the model for child sexual abuse related harms directly involving offenders (i.e.  not viral image sharing) is fundamentally different.
The adversary will spend a significant amount of effort to create nodes in the graph that are sensibly connected and have `normal' activity.
The model envisaged in the paper has a one-to-very-many network effect - effectively `Can I do something with my node that will give me benefit across that highly connected subgraph?'.
We believe that this system will likely be relatively effective (modulo previous points) in these cases.
However, offenders will take a very different approach (a one-to-few relationship) and will therefore look very different to these highly connected nodes in the subgraph structure and related characteristics and therefore the feature vectors.
In child sexual abuse related harms, we expect an adversary to spend effort to convince a small number of neighbours (the people, not the classifiers) that they are `the same'.
This is the precursor to the harm occurring but is also likely to make those neighbour relationships appear `normal' to any classifier as well as the victims involved.
The adversary in this model is not trying to create features that are misclassified; they are trying to convince a vulnerable person that they are what they seem to be, which will directly and indirectly affect some of the characteristics that are harvested for the initial feature vector.
The variance from `normal' in the features that can be seen in graph structure, characteristics and metadata is very likely to be insignificant for competent groomers.
It seems the adversarial model for child sexual abuse harms and the one envisaged in the paper are fundamentally different.

\subsubsection{Proximity of detection to harm}
The abusive account types envisaged in the paper seem to have to exhibit particular characteristics that are statistically significant {\em before} the harm occurs.
For example, a spam account will need to build connections, start to be trusted and so on in order to promulgate the spam when it chooses to do so.
It is entirely reasonable to believe that the system could classify accounts as `spam accounts' as they are in the process of building up to a position of doing harm.
Again, this seems to be completely at odds with child sexual abuse related harms.
Up to the point a groomer becomes known as malicious to the victim, they are, by definition, acting in a manner commensurate with their intended persona.
If they did not, the victim is less likely to believe the groomer's persona and engage in the manner intended, therefore forestalling the harm. 
Even if the system could detect a change in behaviour after the relationship between the two has changed, the significant harm has already happened.
This could be seen as analogous to the compromised account harm, but this is unlikely to be a reasonable analogy.
There seems to be a significant difference in whether the harm occurs before or after a detectable behaviour triggers account removal or other interventions.

There is also a question about the proximity of the harm to the classification event.
For example, for investment scams, harm accrues for each person that receives the scam, believes it, follows the instructions and invests money.
Intuitively, the number of people actually harmed will be significantly less than the number who receive the message.
We are ignoring the harm to Facebook of wasted energy and reputation for the purposes of this point.
In contrast, the exchange of a single image from a victim to an offender leads to harm.
There seems to be a property of the harms akin to hysteresis which will have a direct impact on the harm reduction effect of a classification decision.
This concept needs further thought, but it further demonstrates the difference between child sex abuse harms and platform-related harms.

\subsubsection{Driving subsequent action}
The system is based on a hidden layer neural network model.
It is inherently unexplainable, even more than other AI system designs.
That is, there is no way to explain {\em why} a node was classified as it was, nor {\em how} that conclusion was reached.
However, for the intended use cases, the response of removing accounts or access to specific platform features  is entirely within Facebook's control and policy framework, which does not require any external validation.
If we are to rely on this to drive requests for warrants to allow further investigations, it is unclear how law enforcement could convince a judge or explain to anyone why following a classification through to investigation is a sensible course of action.
Without content, it is unclear how the outcome of this classification system is actionable in a law enforcement context.

It is worthy of note that simply removing content online is insufficient to manage the societal harm that accrues from child sexual abuse. In particular, online artefacts of child sexual abuse are likely to be indicative of serious harms that are occurring or may occur in the near future in other domains. In order to manage the societal harm that manifests online, there must be real world interventions that educate, disincentivise and, in extremis, arrest offenders and safeguard and educate young people and victims. 

\subsubsection{Potential collateral damage}
Commentators are suggesting that techniques such as the one described here could be used to provide law enforcement with tips that would lead to further investigation. 
Given the inherent probabilistic nature of artificial intelligence and machine learning techniques, and the performance characteristics detailed here, it is highly likely that content and behaviour that has nothing to do with child sexual abuse will be classified as such, and the event referred to law enforcement agencies. Furthermore, the triage by law enforcement of this unactionable content slows the overall review process and delays consequential safeguarding.  
For the sake of discussion, assume that law enforcement are able to prosecute an investigation based on such tips, even though we believe that to be extremely unlikely\footnote{Paper-based exercises on content-free CyberTips have yet to suggest a way that they could be used to progress an investigation in the general case.}. 
It is not difficult to imagine police turning up at someone's house to investigate potential child sexual abuse when they have done nothing wrong, but have interacted on social media in a way that has been classified as such. 
Without exculpatory information, for example from content of the conversation, there is no way for either the platforms or law enforcement to discriminate those accounts they should investigate and those that they should not. 
This is likely to lead to a significant number of people being `accused' of doing terrible things by a machine learning algorithm, with effects ranging from removal of their account through to law enforcement intervention. 
We believe that {\bf if implemented as suggested, AI and ML running on metadata alone will be unlikely to assist law enforcement in stopping offenders and safeguarding victims, but will likely lead to a chilling effect on free speech, and infringement on user privacy}.

\subsection{Use of AI in child safety on commodity services}
We do not suggest that AI and machine learning have no place in the detection of child sexual abuse behaviours on large scale platforms.
However, it is clear to the authors that AI techniques will have limited effectiveness when they - and the humans who must provide high quality, ground truth, labelled training data - are limited only to metadata. 
For the majority of the child sexual abuse harm archetypes, the precise characteristics of the harm, and the behaviour of the offenders involved, are unlikely to be amenable to AI and ML techniques alone and we must consider them as part of the systemic response to these issues.

\section{Application of techniques on end-to-end encrypted platforms}\label{section:application}
If there is a moral (and with the imminent advent of the Online Safety Bill, in the UK a legal) responsibility for services to provide a reasonably safe environment for their users, it follows that proportionate responses to threats to these users should be implemented.
Exactly what techniques are proportionate will depend on the details of the service, as we have described elsewhere. However, as we believe we have set out in section~\ref{offenders}, the scale of the threat related to child sexual abuse on at least some services is substantial. Furthermore, we have shown that removal of access to content will render ineffective the most powerful techniques used today to detect some forms of harm.

We will now show some hypothetical applications of some of the tools outlined in section~\ref{section:approaches} as mitigations against some of the harm archetypes detailed in section~\ref{section:offencearchetypes} on hypothetical services with particular characteristics, as detailed in section~\ref{section:servicecharacteristics}.
Recall this paper is intended to stimulate informed debate, not to provide complete solutions.
These examples are intended to be illustrative only; they are not complete, they are not normative and they are not intended to signal a requirement from HMG for these specific technologies.
For the avoidance of doubt, we are seeking to highlight the potential utility of the mitigations most relevant to the harm archetype under consideration. This does not imply that we believe that all of the mitigations need to be implemented in every service, but it hopefully highlights what will be lost in terms of detection if mitigations are not included. Our examples apply to the analogues of the current mainstream platforms; future platforms will almost certainly need different techniques and approaches as features and services evolve.

Our privacy and security impact descriptions are necessarily high level, since these are hypothetical scenarios. An agreed framework for properly assessing the various techniques in various contexts will be necessary and we hope that a first attempt at this will come out of the work being done by the National Research Centre on Privacy, Harm Reduction and Adversarial Influence Online (REPHRAIN) \footnote{http://www.rephrain.ac.uk} for the \href{https://www.safetytechnetwork.org.uk/innovation-challenges/safety-tech-challenge-fund/}{Safety Tech Challenge Fund}\footnote{https://www.safetytechnetwork.org.uk/innovation-challenges/safety-tech-challenge-fund/} in the UK\footnote{We note, for completeness, that this will need interpreting in the context of data protection law and, in the UK, ICO guidance.}. We will not presage the outcome of that work here and so provide only a high level view of the issues.
However, we do note that much of the analysis that has been released publicly by various parties improperly ignores the service context or the counterfactual. Commodity platforms are not academic cryptosystems and changes to them should not be evaluated in that context.

\subsection{Grooming}
We will first consider a commodity, large scale social media service with the usual low-friction (i.e. weak) registration service  where all attributes are merely assertions.
This service provides for peer discovery and initial contact, based on asserted or inferred characteristics including, but not limited to, age (both the age asserted at registration and the age inferred from account activity), sexual identity, interests and other metadata about the user and their social connections that drive the business model of the service\footnote{In this case, we are assuming that the service is driven by advertising and selling harvested data. Other business models may lead to data sets that require different approaches.}.

We will enumerate further the illustrative stages of the grooming process outlined in section~\ref{section:grooming}\ref{section:offenderimageexchange}, and show how different techniques applied during the process could have mitigating effects. We will also try to examine the path where no offender is involved and show how these interactions could be differentiated.

\subsubsection{Offender account creation}
Assume that the offender creates a new account for the purposes of attempting to groom a small number of children. Consider the case where the offender registers the account from a Tor node and then immediately makes 100 contact requests to accounts which all appear to be those of children. Further assume that 95 of those requests are rejected\footnote{This is obviously a cartoon scenario, but suffices to make our point.}.
It would be a reasonably safe assumption that this account is malicious and immediate interventions can be applied to mitigate future harm - from removing the ability to make contact requests through to simply removing the account. From a public safety standpoint, the creator of this account should be directed to resources and charities that offer help to those with a sexual interest in children. Obviously, more advanced heuristics and analytics would be used in any real-world implementation, but we include this to show that some, but certainly not the majority of, potential offender behaviour may be detected early and harm mitigated simply.

In the more likely scenario, the offender will attempt to register a new account in a manner as close to that used by their target persona and would spend time acting like the target persona to build history, commonality and therefore make the persona believable. For obvious reasons, we will not enumerate the specific steps likely to be involved in this phase.
At the end of this phase, the offender's new persona account looks like any normal account relevant to their asserted persona. There may be structural anomalies in the graph around the account node (for example second- and third-order connectivity) but to all intents and purposes, the account is that of a child.
In the authors' opinion, it is unlikely that any AI or machine learning system would reliably detect this account as anomalous at this stage.

It is worth noting here that the those involved in child sexual abuse related activity (excepting viral image sharing) are highly motivated individuals and there is a long history of them evolving tradecraft in order to circumvent protection and detection techniques.
While removal of accounts at this stage will always be a useful tool in deterring offenders early in the escalation of their sexual interest in children, it will always be circumvented by those more advanced offenders.

\subsubsection{Initial Conversations}
Once connected to potential victims, the offender will proceed to converse with them to try to build a rapport. Again, there are two likely paths this could take.

In the first, the offender quickly moves to using language that is obviously intended to coerce the victim. Assume there is a language model running on the client device which has been trained to characterise such conversations with reasonable accuracy.
It would be reasonable to use the language model to warn the potential victim that they are likely to be being groomed and to nudge them towards reporting the conversation for moderation by the platform.
It would not make sense to warn the potential offender in this case and it is worth being explicit that nothing has yet left the user's device unless they have made an explicit decision to report the conversation for moderation.
For the avoidance of doubt, we would expect any moderated content to be used as ground truth (high quality, human labelled) data in any future language model training set, overcoming some of the concerns noted in section~\ref{section:AI}.
Furthermore, it should be obvious that if a human moderator on the platform determines that an offence is likely to have taken place, there exists both metadata and content which can be passed to law enforcement via the relevant child protection NGO.
If there are effective server-side metadata driven heuristics, analytics or AI systems that can provide further context or signals of bad behaviour, these would also feed into the client-local decision  to warn the user.

In the second, the offender is much more careful and spends time building the (false) persona of a peer who is seeking to build a relationship. While language may be sexual, it will be difficult to differentiate at this stage between genuine peer-to-peer sexual messaging and grooming offences.
However, as the interaction is probably approaching the point of potential maximum regret, a language model on the client that holds state that suggests this could be a risky conversation (or otherwise) could be useful, either to provide locally generated warning to the potential victim, or to provide context for decisions being made in the next stage.

\subsubsection{`Proof of intent' through image exchange}
The conversation is now likely to move to image exchange.
It is worth noting here that children's exposure to indecent imagery is not a rare event.
The All Party Parliamentary Group on Social Media's recent report\cite{APPG} cites NSPCC research that shows that $15\%$ of children aged 11 to 18 have been asked to send self-generated images and sexual messages; $7\%$ of 11 to 16 year olds have been asked to share a naked or semi-naked image of themselves; and on average, one child per primary class has been sent or shown a naked or semi-naked image online by an adult.
A recent Ofsted report provides further background on the `normalisation' of these issues with 9 out of 10 children that spoke to the Inspectorate telling them that sexist name calling and being sent unwanted explicit pictures or videos happened `a lot' or `sometimes' \cite{ofsted-report}.

Again, there are two likely paths that could be taken. For the sake of the discussion, assume that any detection is run on both sending and receiving parties.

The first is that a known child sexual abuse image is shared from the offender to the potential victim in order to try to engender trust.
In order to cope with this, we implement a server-server multi-party compute scheme where queries are bound to the published state of NGO child abuse image databases (as described in section~\ref{section:matchingtechniques}).
In this case, the client learns nothing and leaks no meaningful data about the image concerned.
The platform (in our example) learns that the account involved sent an image that is a known child sexual abuse image (but not which one) and the NGOs concerned (one or many of them colluding) learn that a specific image was sent on the platform but not which account sent it.
Assuming the image is detected as known illegal, the client would automatically send the conversation (including the image) for moderation and the conversation would be terminated to stop further harm, with the victim being notified as to why.
We would expect the conversation to be automatically sent for moderation since the probability of illegal activity is very high and the potential for real harm is very proximate to the detection event.
There is then a policy question about whether other accounts in contact with the potential offender accounts should be notified and/or automatically moderated.
As is often the case, there is a balance to be struck in this case and that balance should be the result of democratic debate.
The important point for this paper is that it should be possible, since further children will likely need safeguarding.

The second path is that a sexual image is sent, but not one that is classified as a known illegal image.
A nudity detection capability on the client could be used to help inform users of the fact that they are about to perform a risky act and nudge towards reporting the conversation for moderation if the user is, in retrospect, concerned.
There may be policy choices driven by other metadata and accessible content analysis that could be used here to make more subtle decisions to assist the user, for example making it very hard to send a nude image if the account is believed to belong to a child.
There are further policy considerations as to whether an account that is believed to belong to a child is able to receive images of a sexual nature at all.
This could be an opt-out account policy, for example, which could in itself provide further useful signals as to the likely intent of an encrypted conversation, should a user opt out of this protection during a conversation that raises concerns for other reasons.

\subsubsection{Escalation}
By this point in the conversation, a rapport has been built and images exchanged but there is nothing to suggest that this is an offender-victim communication at the moment. It is possible that this
conversation can easily end up in a harmful situation, regardless of whether the sender is a peer or offender.
However, the potential victim's client now holds state that represents that this is a risky conversation.
Consider yet another language model on the client that is trained to detect coercive, abusive or threatening language.
In this case, it it likely to be appropriate to automatically cease further communication, unless specifically re-enabled by the victim, and to strongly nudge the victim to report the conversation for moderation.
From a socio-technical point of view in designing these interventions, it is important to note that this escalation can happen in a matter of minutes from initial image exchange.

\subsubsection{Privacy and security considerations}\label{subsubsection:privacy-and-security}
First, we consider compromised clients, where any client-side mitigations are removed or otherwise tampered with to reduce their effectiveness. We assume an offender has an arbitrary client that does not participate in safety mechanisms in a manner completely under the control of the offender.
We note that the service characteristics matter here as those that allow web browsers to be first class citizens\footnote{We use this term to mean that browser-based access provides the same features as a thick client. This is important as the implementation of client-side techniques will be very different in these two implementation models.} will be more vulnerable to compromised clients than those that do not.
Furthermore, as software attestation on commodity devices becomes more mainstream, creating compromised apps that  seek to emulate a service's standard app will become ever more detectable by the service.
Regardless, for this harm archetype, it is highly unlikely that the potential victim would have anything other than the client as provided by the service owner.
Therefore {\em in this particular harm archetype} a compromised client does not appear to be of use to the offender.

Next, consider the image classification system used for nudity detection.
In our example, this is likely to be a pre-trained and optimised model, deployed to the client.
This model will be constrained by client characteristics such as processing capacity, power envelope and so on and this will have an effect on the model's accuracy.
We assume that an adversary has to do zero work to obtain and reverse engineer the model from the client software and that it is simple to create adversarial images that do not contain nudity but are classified as such and vice versa.
This does not appear to be a significant issue in this particular example case.

In our primary use case for nudity detection - dissuading or stopping a potential victim from sending a real image of themselves - neither the incentives or likely technical path align.
A potential victim is highly unlikely to go through convoluted technical means to circumvent safety systems in order to send a nude image of themselves, and they are highly unlikely to be using a compromised client in the general case and so will be following the service owner's intended user journey.

From an offender's point of view, there is no point in sending an image that contains no nudity but is classified as such.
This does not help them achieve their goal of coercing the victim and potentially raises any risk score attached to the conversation.
If they can construct a nude image that is not detected, this is useful to the attacker if and only if the image is also convincing to the potential victim.
We believe this is unlikely, but even when an attacker is successful, there remain other defence-in-depth protections in our example mitigation set.

Attacks against the trained model itself, including training set membership inference attacks (discovering that a particular image was part of the training data set) and model inversion attacks (recovering the training data set) do not seem to be particularly harmful in this case, although care must be taken to ensure that the training sets do not contain illegal or identifiable imagery.

In our example, the multiparty compute algorithm used for detection of known child sexual abuse images is a deterministic perceptual hash algorithm like PhotoDNA, rather than a machine learning model.
We assume that the MPC implementation does not require significant algorithm-specific computation on the client, such that determination of the overall detection algorithm is hard.
Here, the main concerns would appear to be knowing what image database is being used for matching, what happens when the client is prevented from communicating with the server side infrastructure\footnote{We would expect any sensible design to make this hard to do from the client, but for the sake of discussion assume this is simple to achieve.} and what information leaks from the client participating in the MPC scheme.

Today, the global NGOs that are responsible for managing the databases of known child sexual abuse content do not maintain a consistent set of images, although those databases are managed and overseen to a consistently high standard.
Assume for the sake of the discussion that all global child protection NGOs that hold such content work together to have a consistent set of content at any given time.
This may be limited to only the highest categories of content and self-reported imagery, or it may be a complete set of all known content.
Assume that there exists a synchronisation protocol between the NGOs such that databases of known bad images are consistent among them at any given point in time.
It is simple, then, for each to publish a cryptographic state that allows confirmation of the content of a given database {\em post priori}, for example the root of a Merkle tree where leaves are some representation of the known-bad images.
It is also possible to then bind any queries to the state of the database (which we assume to be the same as the published state, but design the system to fail if it is not) and other, ancillary information, for example date.
We can further extend this so that concurrence among a number of the global NGOs is needed to allow any multi-party computation to complete.
If we further assume that we can construct some legal mechanism to allow the NGOs to audit each other, we can show that when a query was made against a particular database, the content of that database was such that the cryptographic state could be correctly calculated and that the content had been independently audited (more likely random samples) to contain only known child sexual abuse content.
We believe this sort of construction would address many of the concerns raised around previous potential solutions.
Obviously, these sorts of techniques are not limited to an MPC scheme and more detailed design and analysis would be needed for a real world implementation.

Assume now that such an acceptable client side scanning system exists.
There will be concerns about the ability of offenders to neuter this and therefore any benefit it may bring.
In this particular harm type, the victim party is unlikely to collude with the offender to circumvent safety systems.
Therefore, if protections are run at both ends of the conversation (even on a probabilistic sampled basis) then, even if the offender's client has been compromised, the victim's should still have integrity, providing them with protection.

Significant work will need to be done to fully understand in what circumstances a client side capability leaks information.
At some level, it is axiomatic that it will 'leak' something in the sense of an academic end-to-end cryptosystem - even the single bit of information that reports that a classification decision has been made may be too much for some people.
However, we do not believe that these safety systems should be judged in this manner but rather that a more thoughtful framework for assessing potential client side leaks is necessary.

Attacks against the language models run client side seem to have similar characteristics to those against the image classification model attacks above.

In some situations, it may be preferable to refer content to platform moderators without proximate notification to at least one of the parties, when that moderation was triggered by an automated system.
The case where a moderation is triggered by a positive user action is simple.
In the case where neither party are informed (which must be an automated referral), there will be concerns about future abuse of this capability for surveillance, content discovery and so on.
Since this automated referral is to be driven mainly by content and metadata-based systems on the client, they will be useless as a targeted exceptional access system - such a capability would be quickly discovered through reverse engineering of client code.
Therefore, issues seem to be around the potential detection or surveillance of particular groups of people who consume particular content or behave in particular ways.
We believe it is possible to assuage many of these concerns through providing cryptographic audit functions that allow for transparency.
The design of such systems are beyond the scope of this paper, especially since each situation may require slightly different solutions.

The situation where one party is notified seems more complex, not from a technical point of view but due to the potential for unintended second order social consequences.
It is easy to imagine a situation where two peers are engaging in a conversation that is classified as potential grooming by the system and one of them is warned (and so the other implicitly accused) which could have consequences in the real world for that innocent person.
If this classification is actually a false positive, there is also potential for a chilling effect on free speech as people are falsely accused of engaging in `risky conduct' (or however it is phrased) with all the attendant impacts that could have.
More work is needed to better understand the balances that will inevitably come into play in this scenario.

\subsection{Viral Image Sharing}

While it is hard to succinctly characterise the anatomy of an online viral sharing event, as there is significant variance in the nature of this phenomenon, with child sexual abuse material the motivation is generally misdirected outrage. General online viral events, based around legal memes and popular culture references, typically take place over a period of between 1 and 10 days \cite{anatomy-viral}. Most events, but particularly the longer ones, involve multiple platforms, with the viral media repeatedly criss-crossing between them. This is consistent with what we observe with viral sharing of child sexual abuse material, so this is the high level model we will consider here.

We will not attempt to analyse the totality of a multi-platform, multi-day online viral event here but will instead focus on a single hypothetical platform. It is important to note that there could be a large number of entry points into the platform for the specific piece of child sexual abuse material and that there is a relatively short window of time to respond in order to minimise the spread and hence the harm caused by this activity.

For this scenario we will consider a large scale, commodity messaging service with a typical low friction registration service, where all attributes are merely assertions.
While the service has no peer discoverability, as defined in section~\ref{section:servicecharacteristics}, it does have peer contactability which enables large scale forwards, without first having to establish a trusted connection with each party. For the sake of this discussion, we shall assume that there are no limits on the number of recipients of a message or on the number of times a message can be forwarded, although we note these are useful restrictions in reality. 

At a high-level there are three categories of communication on this platform: sharing in a public space, sharing in a closed group and direct messaging between two users.  Not all of these will be available on every platform but for completeness in this analysis we will assume they are.

Public spaces are those which can be accessed from any user account and are therefore also accessible to the service provider. 
Closed groups are only accessible to accounts that have been granted access by a group administrator and messages are end-to-end encrypted to all group members by default. There are several different architectures for achieving this and the specific details are not pertinent to this analysis, but we do assume that members only have access to group messages sent after they were admitted to the group. Direct messaging is facilitated by peer contactability and all direct messages are end-to-end encrypted.

\subsubsection{Initial detection - public areas}
By definition, the service provider has access to all content that is posted in public areas and so existing server-side techniques can be brought to bear. 
This would enable the media to be detected either prior to delivery or very shortly after delivery, depending on the architecture used. The deterministic known bad image detection algorithms that currently run server-side, such as PhotoDNA, have high efficacy and very low false positive rates, and so it is reasonable to automate the removal of the media once it has been detected, reducing the chance of unwitting recipients viewing the image and limiting the opportunity for further distribution. 

First generation child sexual abuse material will, of course, remain undetected by these checks, as will some variants of known bad imagery. Image classification techniques, for example a combination of nudity detection and age assessment models, could identify such material. These techniques generally have a higher false positive rate and so may only be used to refer a potential case to the platform's moderation team, a process that will naturally take time and so allow for significant spread of the content.

\subsubsection{Initial detection - groups and direct messages}
If imagery is shared outside of the public areas of the platform, server-side approaches will not work since the service does not have access to content. 
Firstly, consider the case where there is no client-side scanning implemented. In this case, we rely solely on user reporting to enable the platform to take action and, as discussed in section~\ref{subsection:user reporting}, the design and and ease of access and use of this functionality is critical. It is reasonable to assume that algorithms running on metadata alone could determine that a particular widespread message is causing negative reactions, but not why. This could be as simple as counting reactions, but we would expect
more complex algorithms to be more effective, for example ones along the lines of those explored in section~\ref{section:AI}.
This could be used to nudge recipients into reporting the content for moderation. However, given that many social media platforms rely on extreme reactions, this is likely to lead to warning fatigue in general and is unlikely to be sustainable as a response. In any case, we would expect a significant timeframe between the initial sending of the child sexual abuse content and action to remove it, based on user reporting, leading to significant spread of the content. 

Now consider the addition of client-side scanning. 
Firstly, we will consider the addition of algorithms that operate solely on the client and do not interact in any way with an external party. The addition of nudity detection and age estimation algorithms to the client could be useful in helping to reduce the warning fatigue that will otherwise be felt by users. In this scenario, users are only nudged to report content for moderation when it is classified as a nude, under 18 image. This could be in addition to or instead of the reaction based nudging. Intuitively, this should increase the chance of a user reporting the content for moderation, but there is little evidence to back up this assertion. In the case of groups, we broadly observe that the groups which are more likely to circulate child sexual abuse material are on average less likely to report it.

Next, we can consider the addition of a server-server MPC deterministic perceptual hash with database integrity protections as outlined in our previous example in order to detect known child sexual abuse images. In this case, detection would cause automated moderation and immediate deletion of known content, reducing the spread and therefore harm. This addition provides no benefit in the case of first generation material being shared. 

\subsubsection{Follow on action}
If the material reported to the platform moderator function is deemed to be child sexual abuse material, the image and associated threads can be removed from the platform and the image referred to the relevant child protection NGO, for example through the NCMEC CyberTip regime. 
The relevant NGO will undertake a further review and, if necessary, add the image to their database of known child abuse images and from then on it is treated as a known child abuse image. 
If first generation imagery is detected, the NGO may  seek to refer the case to law enforcement in the jurisdiction of the originating account for investigation and victim identification and safeguarding. 

In all cases, once an image is determined to be child sexual abuse imagery, the service provider knows from the service metadata the identities of those accounts that shared the content, those that received it and those that re-shared it. 
This means that educational messages can be targeted at the relevant users and, if necessary, action taken against accounts that continue to offend in this way. 

\subsubsection{Analysis}
The characteristics of this harm type - sharing child sexual abuse material in misplaced outrage - does not give rise to significant adversarial concerns. In particular, we would not anticipate the individuals involved in viral image sharing to be attempting to subvert or negate the detection mechanism described above, if they are genuinely acting through misplaced emotion.
Where the client-side nudity and age estimation algorithms are included, there is a risk that adversarial images that do not depict child sexual exploitation but still cause the classifiers to produce a false positive are used to cause warnings to appear for a target set of users or groups. While technically more difficult, the effect of this does not seem to be significantly different to sending similar warnings directly, much like is seen in extortion email scams. There is also the potential for the platform moderation function to be overwhelmed, but this would require large scale reporting, which we do not see today, and a naive workflow management for the moderation function. 

Assuming the nudity and age estimation classifiers are reasonably accurate and that it is likely that at least some users will report child sexual abuse content for moderation when prompted, it is not clear how much extra benefit is accrued by the addition of client-side capability to detect known child sexual exploitation images.

For the case of the the viral image sharing archetype, it is possible that the combination of industry good practice detection mechanisms in public spaces coupled with fairly aggressive limits on group size and forwards and client-side nudity and age estimation algorithms to nudge reporting could be sufficient, without the need for more sophisticated known bad image detection protocols. Given that the focus for this harm type is on prevention and education this could be a reasonable outcome. While the intent for mitigation for this harm archetype is prevention and education, the possibility of detection of first generation imagery in this route should not be discounted and therefore a route to law enforcement for victim identification and safeguarding should be preserved. 

\section{Further Work and Conclusion}

Based on the analysis we have presented, we recommend a number of lines of further investigation which will help to explore the space of potential solutions to the problem of child sexual abuse in communications services, to demonstrate what is practical using technology available today and to build understanding of the risks and advantages of various approaches.

\begin{itemize}
\item{Practical demonstrations of technology\\
Technical safety mitigations cannot exist in isolation, and to understand their properties when integrated into real systems, running on real hardware and used by real people, real demonstrators are invaluable.
Projects such as the Safety Technology Challenge Fund and the similar \href{https://ec.europa.eu/info/funding-tenders/opportunities/portal/screen/opportunities/topic-details/isf-2021-tf1-ag-cyber}{EU programme} help to build a picture of the art of the possible, and yield example systems which provide a concrete basis for grounded conversations exploring the impact of the technologies, as well as some practical implementations of safety technologies which are more mature, and could be integrated into services.}
\item{Technology-assisted safety\\
Whilst some research exists in thie field, furthering our understanding of how users respond to technology-issued nudges to help assist them to be safe in a communication service will allow these to be implemented more effectively and to mitigate potential risks in their adoption.
The understanding needed is not limited to optimising how UI elements may be perceived by users (though this is likely important), but extends to the wider social aspects of suggesting that a child may be likely to be harmed by a communicant.}
\item{Age verification techniques\\
Understanding which users are children and require additional protections may be particularly important for some types of service.
Further improving the accuracy and portability of such techniques would make them more accurate and readily deployable.
Equally important is a detailed exploration of the practical impacts on child users and the system as a whole.}
\item{An evaluation framework\\
In order to facilitate constructive conversations and focus analysis in the most impactful areas, we require an evaluation framework providing a shared context in which to analyse the impacts of safety technology (and indeed other features more generally) on communications systems and their users. 
Models that treat complex real-world services as academic cryptosystems are inadequate for representing the details involved, and a more holistic framework for understanding harms and benefits is needed.
This includes the ability to measure the effectiveness of a range of safety techniques (both social and technical) and how they interact when implemented together.}
\item{Cryptographic audit techniques\\
Practical implementation of auditable interfaces for use in a range of the solutions we have described, providing assurance that safety techniques are used only for their intended purpose.}
\item{Further information on harms and outcomes\\
We have presented in this paper the information that is currently available about the scale of harm, law enforcement actions and safeguarding outcomes in the UK context.
Whilst illustrative of the scale of the problem, further information would allow future system designs to be better targeted to address the harms that occur on each platform.
This might involve a more complete breakdown of the outcomes by category of offence, along with how the activities were identified and on what service they occurred -- so far as is practical given the operational complexities of tracking intelligence and evidence sources to final outcomes.}
\item{Multi-party collation of child sexual abuse image lists\\
We discussed the potential for NGOs to collaborate to achieve common state which could be used to provide increased assurance that a list of child sexual abuse images had not been manipulated. This idea could be further developed to explore the costs and benefits of this sort of collaboration, and, if deemed practical and desirable, to design the protocols and processes that would be involved.}
\item{Impact of service characteristics\\
In section~\ref{section:servicecharacteristics} we set out some of the features of communication services that affect the safety of users.
More detailed study on the effects of these characteristics will faciliate increased accuracy in accounting for them when assessing the risk of harm on each service.}
\item{Multi-party computation\\
As we have highlighted, there are multi-party computation protocols available today which allow for practical implementation of privacy-preserving detection of known child sexual abuse images with what appears to be reasonable resource consumption. Further research into this area of cryptography may improve the efficiency of such protocols, possibly to the extent that detection of first generation images whilst keeping the model private from the client is practical.}
\item{Model inversion attacks and mitigations\\
Whilst it is possible to train models for the detection of child sexual abuse images without resorting to the use of existing known child sexual abuse images to train, it may be possible to improve accuracy by including this material in the modelling process. However, doing so puts this content at risk from potential inversion attacks. Further research on the potency of such attacks, and mitigations that can be implemented to prevent them, would provide greater confidence in assuring that these images could not be recovered from the published model without some measure of prior knowledge of their inclusion.}
\item{In-depth public conversation\\
Services carry responsibility, to some degree, for the activities that they facilitate -- both positive and negative. In order to achieve a balance of opportunities for public good against mitigation of public harm that meets with the expectations of users and of wider society, we welcome further, deeper dialogue on the details concerning the varied policy, service-specific, and technical issues that we have highlighted in this paper.}
\end{itemize}
	
In this paper, we have attempted to lay out the problem of online child sexual abuse and how it manifests on large scale, commodity communication and social media platforms today and how specific platform characteristics affect the type and scale of those offences.
We have enumerated the current mechanisms used to counter this threat and how they will likely be affected when their access to content is removed, for example through end-to-end encryption. Finally, we have considered the types of new approaches that could be employed and provided some examples of how they could be combined to provide child safety on end-to-end encrypted platforms. 

We have tried to be neutral in our presentation and been clear where more work is needed, both in terms of techniques and - more importantly - in the robust and evidence-based evaluation of the efficacy and privacy and security impacts of any solution on a given platform. This is a complex set of problems and there needs to be more work across the community of academics, industry, law enforcement, child protection organisations and even responsible governments in order to come to sensible, pragmatic solutions. 

We do not subscribe to the attempted assessment of safety systems as academic cryptosystems, or the binary presentation of the potential availability of `safe solutions' to this set of problems. The reality is much more subtle and complex, as we have tried to set out. This paper is not a rigorous security analysis, but seeks to show that there exist today ways of countering much of the online child sexual abuse harms, but also to show the scope and scale of the work that remains to be done in this area. As we become more and more dependent on technology to provide even the most basic societal interactions, we must find ways of protecting the most vulnerable users in those circumstances. 
That will require an objective and repeatable evaluation framework, but more importantly a willingness for all parties to engage in rational, evidence-based, scientific discourse to provide the solutions that society requires. 

From our high-level analysis, it is clear that a range of safety features will be needed to address the risks of child sexual abuse in its many forms on communication services, used in combination to mitigate various aspects of the different threats.
Exactly which techniques to implement in what circumstances are nuanced decisions for each service provider, as they consider the best way to provide a safe, secure and private platform for their users.
It is important that the fullest range of techniques is available to counter the threats we have highlighted, and their suitability assessed based on a rigorous and practical assessment of their impacts on the specific systems into which they are integrated.

We have not identified any techniques that are likely to provide as accurate detection of child sexual abuse material as scanning of content, and whilst the privacy considerations that this type of technology raises must not be disregarded, we have presented arguments that suggest that it should be possible to deploy in configurations that mitigate many of the more serious privacy concerns.
AI analysis using metadata available on the server will not (on its own) provide accurate detection of most of the harm archetypes we have described in most probable scenarios.
	
We hope that this paper achieves our intention of informing the important debate around online child sexual abuse and wider user safety. There remains work to be done in order to develop appropriate policies, techniques and evaluation frameworks, but we do believe that solutions exist to many of the issues and problems. We look forward to a continued, informed debate that leads to solutions that can protect both the privacy and security of the general online population, but also protect the vulnerable sub-populations from more specific harms.

\bibliographystyle{abbrvnat}
\bibliography{bibliography}

\end{document}